\documentclass[12pt,bibliography]{article}
\usepackage{hyperref}
\usepackage{epsfig}
\usepackage{amsfonts} 
\usepackage{amsmath} 
\usepackage{color}

\usepackage[utf8]{inputenc}
\usepackage[T1]{fontenc}

\def\half{{1 \over 2}}

\def\tr{{\rm Tr}}

\def\a{\alpha}

\def\Or[#1]{{\text{O}}\left({#1}\right)}
\def\dotl[#1,#2]{\left\langle #1, #2 \right\rangle}
\def\dotlb[#1,#2]{[ #1, #2 ]}
\def\dotp[#1,#2]{(#1) \cdot (#2)}
\def\aff[#1,#2]{\hat{#1}(#2)}
\def\n4sym{{\cal N}=4 SYM}
\def\>{\rangle}
\def\<{\langle}
\def\weight[#1,#2,#3]{\{(#1),#2,#3\}}
\def\ads[#1]{$\text{AdS}_{#1}$}

\newcommand{\ba}{\begin{eqnarray}}
\newcommand{\ea}{\end{eqnarray}}

\usepackage[active]{srcltx}

\newcommand{\be}{\begin{equation}}
\newcommand{\ee}{\end{equation}}  
\newcommand{\bi}{\begin{itemize}}
\newcommand{\ei}{\end{itemize}}

\newcommand{\Ocal}{{\mathcal O}}

\newcommand{\aslash}[1]{\,\,{\raise.15ex\hbox{/}\mkern-12mu #1}}
\newcommand{\bslash}[1]{\,\,{\raise.15ex\hbox{/}\mkern-9mu #1}}

\renewcommand{\bar}{\overline}
\renewcommand{\tilde}{\widetilde}
\renewcommand{\hat}{\widehat}

\newcommand\lrpar{\raise .8ex\hbox{$^\leftrightarrow$} \hspace{-9pt}
\partial}
\newcommand\lpar{\raise .8ex\hbox{$^\leftarrow$} \hspace{-9pt}
\partial}
\newcommand\rpar{\raise .8ex\hbox{$^\rightarrow$} \hspace{-9pt}
\partial}
\newcommand\lrd{\raise .8ex\hbox{$^\leftrightarrow$} \hspace{-9pt}
\nabla}

\newcommand{\gsim}{\lower.7ex\hbox{$\;\stackrel{\textstyle>}{\sim}\;$}}
\newcommand{\lsim}{\lower.7ex\hbox{$\;\stackrel{\textstyle<}{\sim}\;$}}

\let\a=\alpha  \let\g=\gamma  
    \let\k=\kappa
  \let\n=\nu \let\x=\xi \let\r=\rho
\let\s=\sigma     
\let\w=\omega  \let\D=\Delta

\renewcommand{\ba}{\begin{eqnarray}}
\renewcommand{\ea}{\end{eqnarray}}
\newcommand{\bea}{\begin{eqnarray}}
\newcommand{\eea}{\end{eqnarray}}

\def \sha{{\,\amalg\hskip -3.6pt\amalg\,}}

\begin{document}

\begin{titlepage}

\begin{center}
\vspace{1cm}

{\Large \bf The  role of leading twist operators in the\\
\vspace{0.2cm}
Regge and Lorentzian OPE limits}

\vspace{0.8cm}

{\bf Miguel S. Costa${}^a\,$, James Drummond${}^{b,c,d}\,$,  Vasco Gon\c{c}alves${}^a\,$, Jo\~ao Penedones${}^a\,$}

\vspace{1cm}

{\it ${}^a\,$ Centro de F\'{i}sica do Porto \\
Departamento de F\'{i}sica e Astronomia\\
Faculdade de Ci\^encias da Universidade do Porto\\
Rua do Campo Alegre 687,
4169--007 Porto, Portugal}
\\
\vspace{.3cm}
{\it ${}^b\,$CERN, Geneva 23, Switzerland}
\\
\vspace{.3cm}
{\it ${}^c\,$School of Physics and Astronomy, University of Southampton\\
 Highfield, Southampton, SO17 1BJ, U.K.}
 \\
 \vspace{.3cm}
{\it ${}^d\,$LAPTH, CNRS et Universit\'e de Savoie,\\
 F-74941 Annecy-le-Vieux Cedex, France}

\end{center}
\vspace{1cm}

\begin{abstract}
We study two kinematical limits, the Regge limit and the Lorentzian OPE limit, of the four-point function of the stress-tensor multiplet in Super Yang-Mills at weak coupling.
We explain how both kinematical limits are controlled by the leading twist operators.
We use the known expression of the four-point function up to three loops,  to extract  the pomeron residue at next-to-leading order.
Using this data and the known form of pomeron spin up to next-to-leading order, we predict the behaviour of the four-point function in the Regge limit at higher loops. Specifically, we determine the leading log behaviour at any loop order and the next-to-leading log at four loops. 
Finally, we check the consistency of our results with conformal Regge theory. This leads us to predict the behaviour around $J=1$ of the OPE coefficient of the spin $J$ leading twist operator in the OPE of two chiral primary operators.
 
%
\end{abstract}

\bigskip
\bigskip

\end{titlepage}

\tableofcontents
\section{Introduction}

Correlation functions of local operators are the basic observables in a Conformal Field Theory (CFT). Two-point and three-point functions of primary operators are fixed by conformal invariance up to a few dynamical constants (conformal dimensions and structure constants). The four-point functions are more interesting because conformal invariance allows for a general function of two independent  cross ratios,
\be
u = \frac{x_{12}^2 x_{34}^2}{x_{13}^2 x_{24}^2} \,, \qquad v = \frac{x_{14}^2 x_{23}^2}{x_{13}^2 x_{24}^2}\,,
\label{CrossRatios}
\ee
where $x_{ij}^2=(x_i-x_j)^2$ is the square of the distance between points $x_i$ and $x_j$.
This function of $u$ and $v$ is not entirely arbitrary because the operator product expansion (OPE) implies that it has specific power series behaviour in the limit $x_i \to x_j$. Imposing the complete set of these OPE constraints leads to the Conformal Bootstrap program \cite{Polyakov:1974gs} that has experienced a recent revival \cite{Rattazzi:2008pe}.
This is a very promising research avenue but it relies on numerical techniques that are rather opaque for physical intuition. 
Therefore, it is  useful to consider specific limits of the four-point function that reduce the complexity of the OPE constraints and lend themselves to analytical study.

In this paper, we consider two kinematical limits that essentially reduce the four-point function to a function of a single variable.
The first limit, which we call Lorentzian OPE, corresponds to making $x_{12}$ approach the lightcone. In this limit, the OPE is dominated by the operators with lowest twist $\tau$, defined by the difference between conformal dimension $\D$ and spin $J$.
This limit is relevant for phenomenological applications like Deep Inelastic Scattering 
and it has recently been analyzed  in the context of the crossing equations \cite{Fitzpatrick:2012yx, Komargodski:2012ek}.
An important result, derived in \cite{Nachtmann:1973mr}, is the convexity of the leading Regge trajectory $\D(J)$ as a function of the spin $J$.
The leading Regge trajectory is the set of operators of lowest dimension for each  spin $J$, also known has leading-twist operators.

The second kinematical limit we shall consider is the Regge limit, which basically corresponds to performing a  large boost on points $x_1$ and $x_2$ (the precise definition is given in section \ref{sec:ReggeLimit}). Naively, this limit is dominated by the operators with maximal spin. However, since there are operators with arbitrarily large spin, a more careful analysis requires summing up the contributions from an infinite number of operators with increasing spin. This has been done in \cite{CornalbaRegge,Costa:2012cb} using Regge theory methods that involve the analytic continuation of the leading Regge trajectory $\D(J)$ to complex values of spin.

\begin{figure}[t!]
\begin{centering}
\includegraphics[scale=0.25]{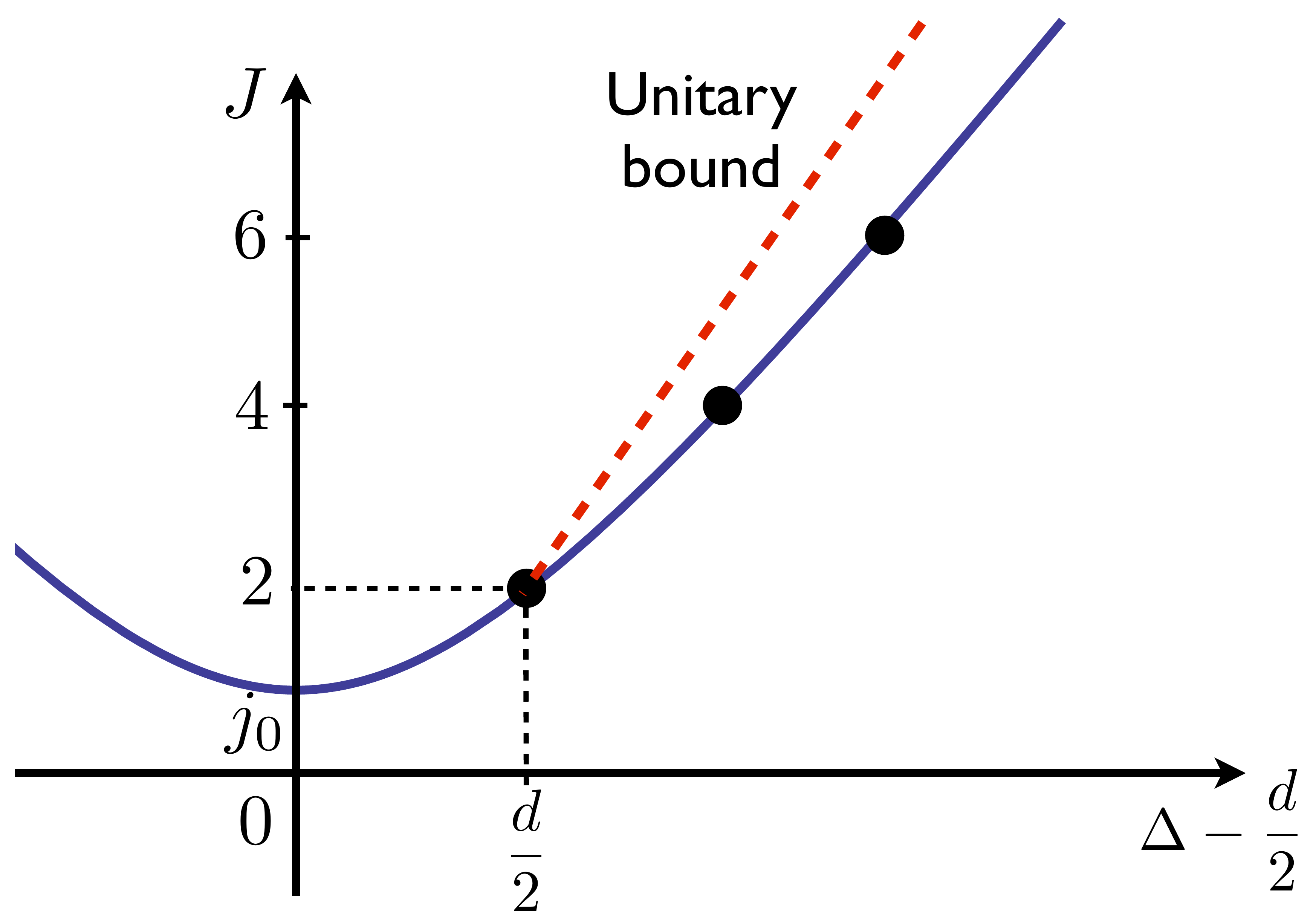}
\par\end{centering}
\caption{\label{fig:DeltaofJ}
Shape of the leading Regge trajectory with vacuum quantum numbers in a CFT.}
\end{figure}
The general expectation for the behaviour of the leading Regge trajectory in a CFT is depicted in figure \ref{fig:DeltaofJ}. We expect the leading-twist operators to play a double role. On the one hand, the operator with minimal twist (usually the stress tensor with protected dimension $\Delta(2)=d$ for a $d$-dimensional CFT) dominates the Lorentzian OPE limit. On the other hand, the Regge limit is controlled by the intercept $j_0$ defined by $\D(j_0)= d/2$, where $\D(J)$ is the analytic continuation of the   dimension of the leading-twist operators of spin $J$.\footnote{Another possibility is that a trajectory of higher-twist operators has an analytic continuation with larger intercept $j_0$ and therefore dominates in the Regge limit.  We think this is unlikely but are not aware of any proof.}

In this paper, we focus on the four-point function of the stress-tensor multiplet in $\mathcal{N}=4$ Super Yang-Mills (SYM), which we review in section \ref{sec:4pt}. 
This is an interesting laboratory because the four-point function is known up to three loops in terms of  (multiple) polylogarithms \cite{Drummond:2013nda} and up to six loops (in the planar limit) as an integral representation \cite{Eden:2012tu}.
The anomalous dimension of leading-twist operators in SYM is also known up to four loops \cite{4loopsTwist2}.
Furthermore, the relevant structure constants or OPE coefficients have  been computed up to three loops \cite{Eden:2012rr}.

\begin{figure}[t!]
\begin{centering}
\includegraphics[scale=0.16]{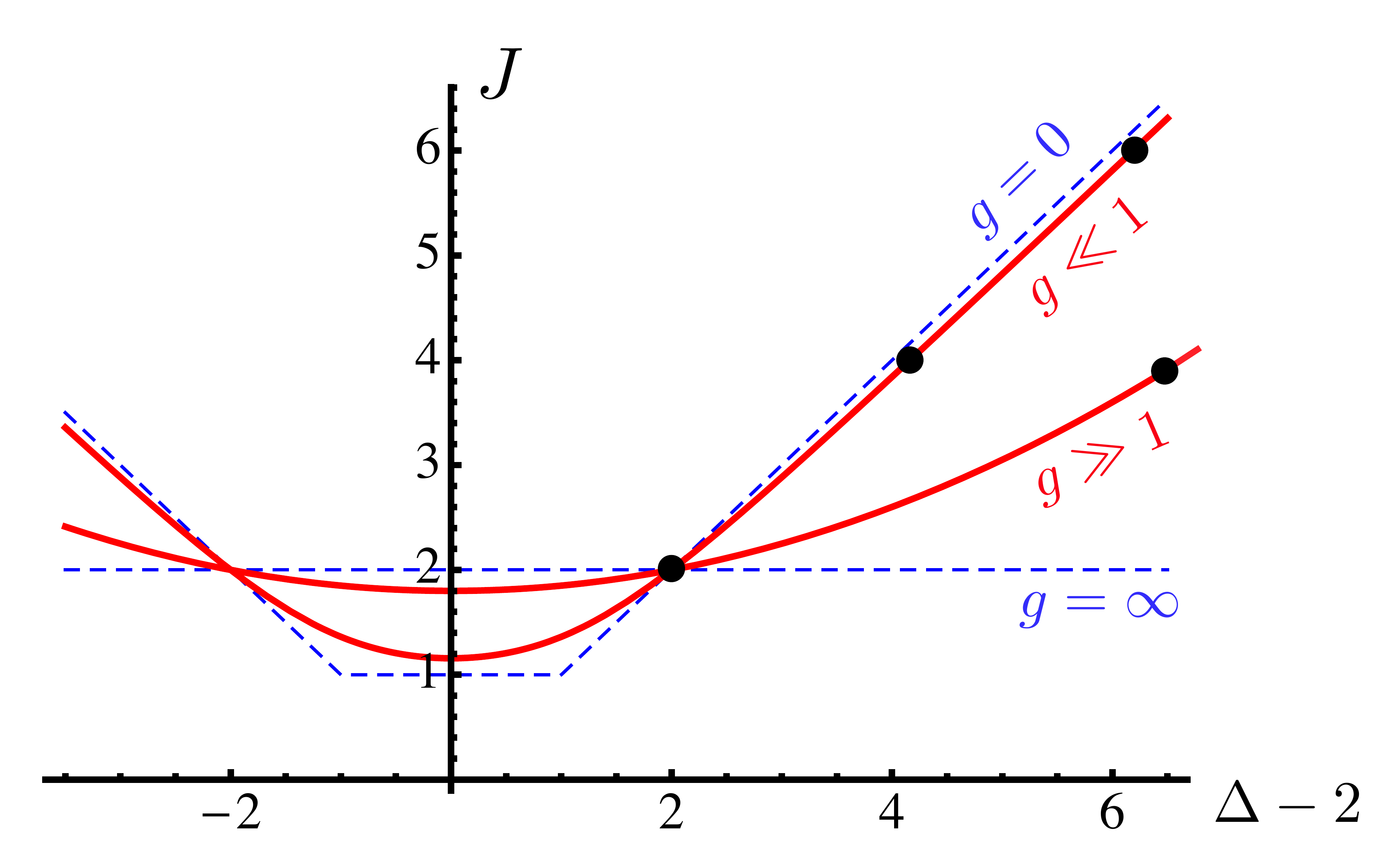}
\par\end{centering}
\caption{\label{fig:DeltaofJweak}
Leading Regge trajectory with vacuum quantum numbers in SYM. At weak coupling $g\to 0$, the trajectory breaks up into three branches. }
\end{figure}

In the weak coupling limit of SYM, the leading Regge trajectory breaks up into three branches  as shown in figure \ref{fig:DeltaofJweak}.
In particular, the horizontal branch that controls the Regge limit and the diagonal branch that controls the Lorentzian OPE limit become almost independent. In particular, the perturbative expansion around each branch of the leading Regge trajectory contains very different information. There is only a small set of consistency conditions that must be satisfied at the meeting point at $\D=3$ and $J=1$  \cite{BPST}. 
From the four-point function point of view, this means that at each order in perturbation theory the Lorentzian OPE limit and the Regge limit provide independent information that can be used to constrain its general form.

At weak coupling, all leading-twist operators have twist equal to two plus a small anomalous dimension. In section \ref{sec:lorOPE}, we review how this leads to logarithms in the Lorentzian OPE limit. In particular, we explain how one can predict the leading logarithms in the Lorentzian OPE limit at higher loop orders using lower-loop information.
In section \ref{sec:ReggeLimit}, we perform a similar analysis of the Regge limit in perturbation theory. In particular, we show that the Regge limit of the four-point function up to three loops is compatible with the predictions of Conformal Regge theory \cite{Costa:2012cb}, and compute the pomeron residue at next-to-leading order. Moreover, we  predict the form of the leading logarithms in the Regge limit at higher loop orders. As explained above, we expect the knowledge of the behaviour of the four-loop four-point function, in the Lorentzian OPE limit and in the Regge limit, to be very useful in constraining a tentative ansatz for the full answer.
In section \ref{sec:CRT}, we explore the   consistency conditions that follow from conformal Regge theory and, in particular, from the fact that the three branches in figure \ref{fig:DeltaofJweak} are a degeneration of a single curve when the coupling constant goes to zero.
Generalizing the approach of \cite{DressingWrapping},  we predict the behaviour of the structure constants between  (the chiral primary component of)  two stress-tensor multiplets   and a leading-twist operator of spin $J$, around $J=1$ to all orders in perturbation theory. 

\section{Four-point function of stress-energy multiplets}
\label{sec:4pt}

Our central object of study is the four-point function of stress-energy multiplets in $\mathcal{N}=4$ SYM theory. The super primary operator for the supermultiplet is constructed by taking the trace of a particular bilinear combination of scalars fields,
\be
\mathcal{O}(x,y) = y^I y^J \tr \bigl(\phi_I (x) \phi_J(x) \bigr)\,,
\label{Oxy}
\ee
where $y_I$ denote a set of auxiliary complex variables transforming in the vector representation of $SO(6)$ and obeying $y^I y^I =0$. All other operators in the supermultiplet can be obtained from (\ref{Oxy}) by applying supersymmetry transformations. 

The stress-energy multiplet is $\tfrac{1}{2}$-BPS. This implies that the four-point functions of any operators in the multiplet are uniquely determined in terms of the four-point functions of the primary operators $\mathcal{O}(x,y)$. Therefore we will focus on the four-point function
\be
G(1,2,3,4) = \langle \mathcal{O}(x_1,y_1) \mathcal{O}(x_2,y_2) \mathcal{O}(x_3,y_3) \mathcal{O}(x_4,y_4) \rangle\,.
\label{4pt}
\ee
The correlation function $G$ is a homogeneous polynomial of degree 2 in each of the $y_i$ variables. There are six such polynomials reflecting the fact that there are precisely six different $SO(6)$ channels in the decomposition of the four-point function.

Superconformal symmetry imposes strong constraints on the form of the correlation function $G$. In fact it takes the form
\be
G(1,2,3,4) = G^{(0)}(1,2,3,4) + R(1,2,3,4) \frac{F(u,v)}{x_{13}^2 x_{24}^2}\,.
\label{G1234decomp}
\ee
The first term above, $G^{(0)}(1,2,3,4)$, is a rational function of the $x_i$ and can be identified with the leading-order contribution to the correlation function in perturbation theory,
\begin{align}
G^{(0)}(1,2,3,4)  =& \ \frac{(N^2-1)^2}{4 (4 \pi^2)^4}\biggl(\frac{y_{12}^4 y_{34}^4}{x_{12}^4 x_{34}^4} + \frac{y_{13}^4 y_{24}^4}{x_{13}^4 x_{24}^4}  +\frac{y_{14}^4 y_{23}^4}{x_{14}^4 x_{23}^4}  \biggr) \notag \\
&\ +\frac{N^2-1}{(4 \pi^2)^4} \biggl(\frac{y_{12}^2 y_{23}^2 y_{34}^2 y_{41}^2}{x_{12}^2 x_{23}^2 x_{34}^2 x_{41}^2} + \frac{y_{12}^2 y_{24}^2 y_{43}^2 y_{31}^2}{x_{12}^2 x_{24}^2 x_{43}^2 x_{31}^2}  + \frac{y_{13}^2 y_{32}^2 y_{24}^2 y_{41}^2}{x_{13}^2 x_{32}^2 x_{24}^2 x_{41}^2}  \biggr)\,.
\end{align}
Here we have used the notation $y_{ij}^2 = y_i \cdot y_j = y_i^I y_j^I$.

The second term in (\ref{G1234decomp}) consists of a coupling-independent factor, $R(1,2,3,4)$, and a $y$-independent factor,
 $F(u,v)$. The factor $R$ is given explicitly by
\begin{align}
R(1,2,3,4)  = \frac{2(N^2-1)}{(4\pi^2)^4}\biggl( & \frac{y_{12}^2 y_{23}^2 y_{34}^2 y_{41}^2}{x_{12}^2 x_{23}^2 x_{34}^2 x_{41}^2} \bigl( x_{13}^2 x_{24}^2 - x_{12}^2 x_{34}^2 - x_{14}^2 x_{23}^2\bigr) \notag \\
+& \frac{y_{12}^2 y_{24}^2 y_{43}^2 y_{31}^2}{x_{12}^2 x_{24}^2 x_{43}^2 x_{31}^2} \bigl( x_{14}^2 x_{23}^2 - x_{12}^2 x_{34}^2 - x_{13}^2 x_{24
}^2\bigr) \notag \\
+& \frac{y_{13}^2 y_{32}^2 y_{24}^2 y_{41}^2}{x_{13}^2 x_{32}^2 x_{24}^2 x_{41}^2} \bigl( x_{12}^2 x_{34}^2 - x_{13}^2 x_{24}^2 - x_{14}^2 x_{23
}^2\bigr) \notag \\
+& \frac{y_{12}^4 y_{34}^4}{x_{12}^2 x_{34}^2} + \frac{y_{13}^4 y_{24}^4}{x_{13}^2 x_{24}^2} + \frac{y_{14}^4 y_{23}^4}{x_{14}^2 x_{23}^2} \biggr)\,.
\end{align}
The factor $F(u,v)$ is a function of the two conformal cross-ratios,
\be
u = \frac{x_{12}^2 x_{34}^2}{x_{13}^2 x_{24}^2} \,, \qquad v = \frac{x_{14}^2 x_{23}^2}{x_{13}^2 x_{24}^2}\,,
\label{CrossRatios}
\ee
as well as, implicitly, the gauge coupling and gauge group. Since the leading order contributions to the correlator are captured by $G^{(0)}(1,2,3,4)$, the function $F$ actually receives contributions in perturbation theory from one loop onwards,
\be
F(u,v) = \sum_{l=1}^{\infty} g^{2l} F^{(l)}(u,v)\,.
\label{ExpansionF}
\ee
Here we have chosen to express the perturbative expansion in terms of the 't Hooft coupling,
\be
g^2 = \frac{g_{\rm YM}^2 N}{16 \pi^2}\,.
\ee
Restricting to the gauge group $SU(N)$, the functions $F^{(l)}$ also have an implicit dependence on the number of colours $N$. In fact the first genuine $1/N^2$ corrections arise at four loops \cite{Eden:2012tu}.
The fact that the coupling dependence factorises from the $y$-dependence is a non-trivial consequence of superconformal symmetry, and has been referred to as `partial non-renormalisation' \cite{Eden:2000bk}. As made explicit in (\ref{G1234decomp}), it follows that the full correlation function is determined by its tree-level part and a single function of two variables, $F(u,v)$.

\subsection{Weak coupling expansion}

The function $F(u,v)$ has an interesting structure in perturbation theory. Up to three loops it can be expressed in the following form,
\begin{align}
F(u,v) = & \quad \frac{f(z,\bar z)}{z-\bar z} \notag \\
&+ \frac{1}{(z-\bar z)^2}\Bigl( g(z,\bar z) + u g\bigl(\tfrac{1}{z}, \tfrac{1}{\bar z}\bigr) + v g\bigl(\tfrac{1}{1-z},\tfrac{1}{1-\bar z} \bigr) \Bigr) \notag \\
&+ \frac{1}{z - \bar z}\biggl( \frac{1+ u}{1- u} h(z , \bar z) - \frac{1+v}{1- v} h(1-z, 1-\bar z) +\frac{u + v}{u - v} h \bigl(1-\tfrac{1}{z} , 1- \tfrac{1}{\bar z}\bigr)\biggr)\,.
\label{Fuvto3loops}
\end{align}
Here we have employed the variables $z,\bar z$ which are related to the cross-ratios $u,v$ as follows,
\be
u= z \bar z,\qquad v=(1-z)(1-\bar z)\,.
\label{uvzzbar}
\ee
The functions $f$ and $h$ are antisymmetric in the exchange of $z$ and $\bar z$ while $g$ is symmetric. It may seem that we have taken a step backwards in expressing a single two-variable function $F$ in terms of three two-variable functions $f$, $g$  and $h$. However the functions $f$, $g$ and $h$ all have the property that their perturbation expansion is expressed in terms of pure transcendental functions of degree equal to twice the loop order. All three functions are single-valued in the Euclidean region where $z$ and $\bar z$ are related by complex conjugation. 

We write the perturbative expansion of the functions $f,g$ and $h$ as follows\footnote{We hope that the context is sufficient to distinguish between the 't Hooft coupling $g^2$ and the function $g(z,\bar z)$.},
\be
f(z,\bar z) = \sum_l g^{2l} f^{(l)}(z,\bar z)\,, \qquad g(z,\bar z) = \sum_l g^{2l} g^{(l)}(z,\bar z)\,, \qquad h(z,\bar z) = \sum_l g^{2l} h^{(l)}(z,\bar z)\,.
\ee
Up to two loops, the functions $f$ and $g$ can be expressed in terms of the following single-valued functions appearing in the ladder integrals,
\be
\phi^{(l)}(z,\bar z) = \sum_{r=0}^l \frac{(-1)^r (2l-r)!}{r! (l-r)! l!} \log^r (z\bar z)\bigl({\rm Li}_{2l-r}(z) - {\rm Li}_{2l-r}(\bar z)\bigr)\,.
\ee
Explicitly we have at one loop \cite{Eden:1998hh,Eden:1999kh},
\begin{align}
f^{(1)}(z,\bar z) = - \phi^{(1)}(z,\bar z)\,, \quad g^{(1)} = 0\,, \quad h^{(1)} =0\,.
\end{align}
At two loops we have \cite{Eden:2000mv}
\begin{align}
f^{(2)}(z,\bar z) &= 2\Bigl[\phi^{(2)}(z,\bar z) + \phi^{(2)}\bigl(1-\tfrac{1}{z},1-\tfrac{1}{\bar z}\bigr) + \phi^{(2)}\bigl(\tfrac{1}{1-z},\tfrac{1}{1-\bar z}\bigr)\Bigr]\,. \notag \\
g^{(2)}(z,\bar z) &= \frac{1}{2}\phi^{(1)}(z,\bar z)^2\,. \notag \\
h^{(2)}(z,\bar z) &= 0\,.
\label{fgh2loops}
\end{align}
At three loops, in addition to ladder integrals, there is the `tennis-court' integral (which is in fact identical to the three-loop ladder \cite{Drummond:2006rz}) as well as two genuinely new integrals, known as Easy and Hard \cite{Eden:2011we}. These new three-loop integrals were evaluated in \cite{Drummond:2013nda}. The result is that at three loops we have
\begin{align}
\label{fgh3loops}
f^{(3)}(z,\bar z) =  &-  6\Bigl[\phi^{(3)}(z,\bar z) + \phi^{(3)}\bigl(1-\tfrac{1}{z},1-\tfrac{1}{\bar z}\bigr) + \phi^{(3)}\bigl(\tfrac{1}{1-z},\tfrac{1}{1-\bar z}\bigr)\Bigr] \notag \\
&-2\Bigl[E(1-z,1-\bar z) - E\bigl(1-\tfrac{1}{z},1-\tfrac{1}{\bar z}\bigr) - E(z,\bar z) \notag \\
&\qquad + E\bigl(\tfrac{z}{z-1},\tfrac{\bar z}{\bar z -1}\bigr) + E\bigl(\tfrac{1}{z},\tfrac{1}{\bar z}\bigr) - E\bigl(\tfrac{1}{1-z},\tfrac{1}{1-\bar z}\bigr)\Bigr]\,, \notag \\
g^{(3)}(z,\bar z) = &- 2 \phi^{(1)}(z,\bar z) \phi^{(2)}\bigl(1-\tfrac{1}{z},1-\tfrac{1}{\bar z}\bigr) - H^{(a)}(z,\bar z) - H^{(a)}(1-z,1-\bar z)\,, \notag \\
h^{(3)}(z,\bar z) = & - 2\Bigl[ E(1-z,1-\bar z) + E(1-\tfrac{1}{z}, 1- \tfrac{1}{\bar z})\Bigr] - H^{(b)}(1-z,1-\bar z) \,.
\end{align}
The function $E$ arises in the Easy integral while the functions $H^{(a)}$ and $H^{(b)}$ arise in the Hard integral. Their explicit form can be found in \cite{Drummond:2013nda}. The functions $E$ and $H^{(b)}$ can be expressed in terms of single-variable harmonic polyogarithms. Only the function $H^{(a)}$ involves genuine multiple polylogarithms.

\subsection{Conformal block expansion}

The conformal block expansion of (\ref{4pt}) was analyzed in detail in
\cite{Dolan:2001tt}. Here, we summarize the main points relevant for our analysis.

To understand the operators that are exchanged in the OPE,
recall that the 
tensor product of two 20's (symmetric traceless two-index tensor of $SO(6)$) decomposes in irreducible representations as follows
\begin{align}
20\otimes20=1\oplus15\oplus20\oplus84\oplus105\oplus175\,.
\label{representations}
\end{align} 
The four-point function can then be expanded in R-charge sectors,\footnote{We normalize the reduced correlators $A_R$ such that the identity operator contributes 1 to $A_1$.}
\be
G(1,2,3,4)= \frac{(N^2-1)^2}{4 (4 \pi^2)^4}
\frac{20}{x_{12}^4x_{34}^4}
\sum_R P_R(y_1,y_2,y_3,y_4) \,A_R(u,v) 
\ee
where $R=1, 15, 20, 84, 105, 175$ denotes the  irreducible representations of $SO(6)$ and $P_R$ are projectors given in appendix \ref{App:Projectors}.
Using the explicit form of the projectors, we conclude that  
 \begin{align}
A_1&=1+ \frac{u^2(1+v^2)}{20v^2} +
   \frac{  u \big(u+10 (v+1)\big)}{15  v(N^2-1)} +\frac{  2u \big(u^2-8 u (v+1)+10 \big(v
   (v+4)+1\big)\big) F(u,v)}{15 v (N^2-1)}\,,
   \nonumber\\
   A_{20}&=
    \frac{u^2(1+v^2 )}{20v^2}
  +\frac{  u \big(u+10 (v+1)\big)}{30  v(N^2-1)}
    +\frac{  u \big(u^2-5 u (v+1)+10 (v-1)^2\big) F(u,v)}{15v(N^2-1)} \,,
\label{A1A20}
\end{align}
and similar expressions are given in appendix \ref{App:Projectors}
for the other  $A_{R}$.

Each $R$-charge sector can then be expanded in conformal blocks \cite{Ferrara:1974nf}
\begin{align} 
 A_R(u,v) = \sum_{\Delta,J} a^R_{\Delta,J} \,G_{\Delta,J}(u,v)\,,
\label{CBexpansion}
\end{align}
where  the constant $a_{\Delta,J}^R$ is given by the product of the OPE coefficients that  appear in the 
three-point function of two external operators and the primary operator of dimension $\Delta$, spin $J$ and $R$-charge 
representation $R$.\footnote{If there is degeneracy of operators, then $a_{\Delta,J}^R$ will be given by a sum over products of OPE coefficients.}
The conformal block $G_{\Delta,J}$ was written in \cite{Dolan:2000ut} in terms of hypergeometric $\, _2F_1$ functions for the present case of $d=4$ spacetime dimensions.

The four-point function is a rather involved function of two variables ($u$ and $v$). It is convenient to consider kinematical limits where it simplifies significantly. In the next sections, we consider two such limits that reduce complexity  to a function of a single variable.

\section{Lorentzian OPE limit}
\label{sec:lorOPE}

The Lorentzian OPE limit is defined by the limit where  $x_{12}$ becomes light-like.
In terms of the conformal invariant cross-ratios, this means that $u\rightarrow 0$ with $v$ fixed. 
\subsection{Weak coupling expansion}

It is instructive to consider the Lorentzian OPE limit of each perturbative contribution to the four-point function (\ref{4pt}).
The general structure is given by
\be
F^{(l)}(u,v)=   \sum_{k=0}^l
 (\log u)^{l-k} \theta_{k}^{(l)}(v) + \Ocal(u)
 \label{FOPElimit}
\ee
where $k=0$ is called leading log, $k=1$ next-to-leading log, etc. Recalling the form (\ref{uvzzbar}) of the cross-ratios in terms of $z$ and $\bar z$, we see that the Lorentzian OPE limit can be implemented by taking $\bar z \rightarrow 0$, leaving $z$ fixed. In this limit the variable $v$ becomes simply $1-z$, and it is convenient to express the results for $\theta_{k}^{(l)}(v)$ in terms of $z$. 

Taking the limit on the form of $F(u,v)$ up to three loops, given in (\ref{Fuvto3loops}), we find that the rational prefactors simplify to be either $1/z$ or $1/z^2$. This means that up to three loops the functions $\theta^{(l)}_k(v)$ are always of the form 
\be
\theta^{(l)}_k(v) = \frac{1}{z}\, \kappa^{(l)}_{k }(z) + \frac{1}{z^2}\, \tilde{\kappa}^{(l)}_{k}(z)
\label{kappa}
\ee
for pure transcendental functions $\kappa^{(l)}_{k }$ and $\tilde{\kappa}^{(l)}_{k}$. 

In order to present explicit results for the limits considered in this paper, it is useful to introduce the family of harmonic polylogarithms (or HPLs) \cite{Remiddi:1999ew}. We will need the harmonic polylogarithms whose weight vectors $w$ are composed of the letters 0 or 1. For a string of $n$ zeros we define $H_{0_n}(z) = \tfrac{1}{n!}\log^n z$. The remaining functions are then defined as follows,
\be
H_{0, w}(z) = \int_0^z \frac{dt}{t} H_{w}(t)\,, \qquad H_{1,w}(z) = \int_0^z \frac{dt}{1-t} H_w(t)\,.
\ee
The classical polylogarithms functions are the harmonic polylogarithms whose weight vectors are are string of zeros followed by a single one, ${\rm Li}_{n}(z) = H_{0_{n-1}, 1}(z)$. As is common in the literature we will employ the shorthand notation whereby a string of $(k-1)$ zeros followed by a 1 is contracted to the label $k$, for example ${\rm Li}_4(z) = H_{0,0,0,1}(z)= H_4(z)$ or  $H_{0,1,0,0,1}(z) = H_{2,3}(z)$.

Expressing the OPE limit in terms of the harmonic polylogarithms we find at one loop,
\be
F^{(1)}(u,v) = \log u \, \frac{1}{z} \, H_1 - \frac{2}{z} H_2  + O(u)\,.
\ee
where we have left the argument $z$ of the HPLs implicit. 
At two loops we have
\begin{align}
&F^{(2)}(u,v)  =  \,\log^2 u \left( \frac{2}{z}\,H_2  + \frac{2}{z^2} \,H_{1,1} \right) \notag \\
& +\log u \left( - \frac{2}{z}\bigl(6 H_3 + H_{1,2} - H_{2,1} \bigr)-\frac{4}{z^2}\bigl(H_{1,2} +2 H_{2,1} \bigr)\right) \notag \\
&  +\frac{2}{z}\bigl(12 H_{4} + 3H_{1, 3} + H_{2, 2} - 4 H_{3, 1} + 2 H_{1, 1, 2} - 2 H_{1, 2, 1} + 6 H_{1} \zeta_3\bigr) 
+\frac{8}{z^2}\bigl(H_{2,2} + 2 H_{3,1} \bigr)\notag \\
& +O(u)\,.
\end{align}
Moving to three loops we quote here the result for the coefficient of $\log^3 u$, 
\be
\theta^{(3)}_0(1-z) = -\frac{4}{3z}\bigl(H_{1,2} - 2 H_3\bigr) + \frac{4}{3z^2}\bigl(4H_{1,2} + 2 H_{2,1} + 3 H_{1,1,1}\bigr)\,
\ee
and the coefficient of $\log^2u$,
\begin{align}
\theta_1^{(3)}(1-z) = &-\frac{4}{z} \bigl(8 H_{4} - H_{1, 3} - H_{2, 2} - 
   2 H_{3, 1} + H_{1, 2, 1} - H_{2, 1, 1}\bigr) \notag \\
   &-\frac{4}{z^2} \bigl(8 H_{1, 3} + 8 H_{2, 2} + 8 H_{3, 1} + 
   4 H_{1, 1, 2} + 3 H_{1, 2, 1} + 5 H_{2, 1, 1}\bigr)\,.
\end{align}

\subsection{Anomalous dimensions and OPE coefficients 
}
\label{sec:AnomDim}

In the Lorentzian OPE limit,   the conformal block simplifies to
\begin{align}
G_{\Delta,J} (u,v)=u^{\frac{\Delta-J}{2}}\left(\frac{v-1}{2}\right)^J  F ( \Delta+J,1-v) \Big[
1+O(u)\Big]\ ,
\label{CBLim}
\end{align} 
with
\be
 F\!\left(x,z\right)\equiv   \,\!_2F_1\!\left(\frac{x}{2},\frac{x}{2},x,z\right) .
\ee
We conclude that the operators that dominate in the Lorentzian OPE are those with lowest twist $\D-J$. 
The identity operator has  twist zero and contributes 1 to $A_1$. 
In SYM at finite coupling constant, the next minimal twist is 2 and corresponds to a small number of protected operators like conserved currents and the stress-energy tensor.
However, in the weak coupling limit, an infinite tower of operators (with any spin $J$) have twist 2. We shall focus on their contribution.

Consider first the $R$-charge sector 20, using the expansion (\ref{FOPElimit}) for the function $F(u,v)$  in (\ref{A1A20}), we obtain
\be
A_{20}=\frac{  u }{3v(N^2-1)}
\left[(1+v) +2(1-v)^2 \sum_{l=1}^\infty
a^l\sum_{k=0}^l
 (\log u)^{l-k} \theta_{k}^{(l)}(v)
\right]
+O(u^2)\,.
\label{A20WCE}
\ee
%
In the 20 channel there is only one twist two primary operator for each  even  spin $J$ that can be exchanged. This is usually written as
$\tr\left(ZD^JZ\right)$ where $D$ is a lightcone derivative and $Z$ is a complex scalar field of SYM.
Therefore, we can write
\be
A_{20}=
\sum_{J=0 \atop {\rm even} }^\infty
a_J \, u^{1+\frac{\g(J) }{2}}\left(\frac{v-1}{2}\right)^J  F ( 2 +2J+\g(J),1-v)  +O(u^2) 
\label{A20CBE}
\ee
 where $\g(J)$ are the anomalous dimensions of the twist operators in the $R$-charge representation 20.
Equating (\ref{A20WCE}) and  (\ref{A20CBE}), it is possible to read the anomalous dimension and OPE coefficients in an efficient way   \cite{Dolan:2004iy,Eden:2012rr}. 
In particular,  \cite{Eden:2012rr} proposed a nice structure for the coefficients  
\be  
a_{J}=\frac{2}{3(N^2-1)}
\frac{2^J\left(1+\frac{\g(J)}{2}\right)_{\!J}^{\ 2}}{\big(1+\g(J)\big)_{2J}}\sum_{n=0}g^{2n}a_n(J) \ ,
\label{eq:aJ}
\ee
using the results up to three loops.
Here we propose to express the pre-factor in terms of 
the Pochammer symbol $\left(a\right)_x$, which simplifies the expression in \cite{Eden:2012rr}. In this form
the functions $a_n(J)$ can be written in terms of harmonic sums of transcendentality $2n$. 
 
Regarding the anomalous dimension of these operators, integrability methods gave a spectacular boost to perturbative information, with explicit expressions 
known up to five loops. 
More recently, several works found perturbative data on OPE coefficients involving twist two operators 
\cite{Eden:2012rr,Plefka:2012rd,Kazakov:2012ar,Alday:2013cwa}.
Writing the perturbative expansion in the simple form, 
\be
\gamma(J)=\sum_{n=1}g^{2n}\gamma_n(J) \ ,
\label{eq:AnomalousDim}
\ee
it is observed that at each order in perturbation theory the functions $\gamma_n$ have definite transcendentality (equal to $2n-1$). 
The explicit expressions for $\g_n$ and $a_n$ up to $n=3$ (three-loops) are given in appendix \ref{App:GammaOPEcoefs}.

In some special regimes, for example in the  large spin limit $J\rightarrow \infty$, it was possible to determine all-loop information for the anomalous dimension \cite{Beisert:2006ez,Basso:2007wd}. For OPE coefficients an all-loop formulae, in the large $J$ limit, was conjectured in  \cite{Alday:2013cwa} (in fact this also implies an all-loop structure for the four point function).

Let us now consider the $R$-charge singlet channel of the four-point function.
In this case, the conformal block expansion is more complicated because there are   three twist-two operators that can be exchanged for each even spin $J$. 
At tree level they can be written as linear combinations of  the following operators,
\begin{align}
{\rm tr} \left( F_{\mu \nu_1} D_{\nu_2} \dots D_{\nu_{J-1}} F_{\nu_J}^{\ \ \mu} \right),\ \ 
{\rm tr} \left(  \phi_{AB} D_{\nu_1} \dots D_{\nu_{J}}\phi^{AB}  \right),\ \ 
{\rm tr} \left(   \bar{\psi}_{A} D_{\nu_1} \dots D_{\nu_{J-1}} \Gamma_{\mu_J} \psi^{A} \right).
\label{zeroorderstates}
\end{align}
At one loop only specific combinations of these operators form primary operators defining three different Regge trajectories;
their expressions are given in appendix E of \cite{Costa:2012cb}. 
The anomalous dimensions of all the above twist-two operators are related by supersymmetry. Defining the anomalous dimension
of the spin $J$ operator in the $20$ channel as 
$\gamma=\gamma(J)$, then the three twist-two operators of spin $J$ in the singlet channel have anomalous dimensions given by
\begin{align}
\gamma(J+2),  \ \ \ \ \ \ \ \ \ \ \ \ \ \ \ \ \  \ \gamma(J), \ \ \  \  \ \ \ \ \ \ \ \ \ \ \ \ \ \gamma(J-2).
\end{align}
The anomalous dimension for operators in the leading Regge trajectory, which contains the energy-momentum tensor, is the one with
$\gamma=\gamma(J-2)$.

Thus, it is not surprising that OPE coefficients between protected scalars from the 20'  representation and twist-two operators in the 
representations (\ref{representations}) can also be related by supersymmetry. At the level of the four-point function, and for twist-two operators
in the singlet and $20$ representations,  this is encoded in (\ref{A1A20}) which can be used to relate $A_1$ and $A_{20}$. 
In the $u\to 0$ limit this relation becomes  
\be
A_{1}-1+\frac{4 u (1+v)}{ (N^2-1) (1-v)^2}=
2\,\frac{1+4v+v^2}{(1-v)^2}A_{20} +O(u^2) \ .
\label{RelationA1A20}
\ee 
On the other hand, the singlet channel four-point function can also be expanded in conformal blocks
\begin{align}
A_{1}=&\ 1+
u\sum_{J=2 \atop {\rm even} }^\infty
b_J \, u^{\frac{\g(J-2) }{2}}\left(\frac{v-1}{2}\right)^J  F \big(2+  2J+\g(J-2),1-v\big) \nonumber\\&
+u\sum_{J=2 \atop {\rm even} }^\infty
b_J' \, u^{\frac{\g(J) }{2}}\left(\frac{v-1}{2}\right)^J  F \big(2+  2J+\g(J),1-v\big)  \label{A1CBE}
\\&
+
u\sum_{J=0 \atop {\rm even} }^\infty
b_J'' \, u^{\frac{\g(J+2) }{2}}\left(\frac{v-1}{2}\right)^J  F \big(2+  2J+\g(J+2),1-v\big)   
  +O(u^2)\,,
  \nonumber
\end{align}
where the 1 corresponds to the contribution of the identity and we denoted by $b_J$, $b_J'$ and $b_J''$ the product of OPE coefficients associated with the three trajectories of twist-two operators that exist in the singlet sector.
Using the relation (\ref{A1A20}) between the two channels $A_1$ and $A_{20}$, and their conformal block expansions (\ref{A1CBE}) and (\ref{A20CBE}), we conclude that
\be 
b_2  \left(\frac{v-1}{2}\right)^2
  F (6,1-v)  
  +\frac{4  (1+v)}{ (N^2-1) (1-v)^2}=
2\,\frac{1+4v+v^2}{(1-v)^2}\,a_0\, 
 F (2,1-v) 
 \label{orderu}
\ee
and
\begin{align}
&12\left(\frac{1}{z^2}-\frac{1}{z}+\frac{1}{6}\right)\sum_{J=2}^\infty a_J\,
u^{\gamma/2}\,\frac{z^{J}}{2^J}\,  F\!\left(1+J+\frac{\gamma}{2},z\right)=
\label{RelationOPEs}
\\
&\sum_{J=2}^\infty\frac{z^Ju^{\gamma/2}}{2^J}\!\left(b_{J+2} \,\frac{z^2}{4}\, F\!\left(3+J+\frac{\gamma}{2},z\right)
+b_J'\, F\!\left(1+J+\frac{\gamma}{2},z\right)+
b_{J-2}''\,\frac{4}{z^2}\, F\!\left(J-1+\frac{\gamma}{2},z\right)\!\right),
\nonumber
\end{align} 
where we used $v=1-z$ and shifted the summation variable $J$ so that the function $\gamma$ has argument always given by $J$.
Equation (\ref{orderu}) follows from   the twist-two contributions to (\ref{A1A20}) with no anomalous dimensions (like the energy-momentum tensor), and leads to 
\be
a_0 = \frac{2}{3 (N^2-1) }\ ,\ \ \ \ \ \ \ 
b_2 = \frac{8}{45 (N^2-1) }\ .
\label{a0b2}
\ee
Equation (\ref{RelationOPEs}) encodes the contributions of all other twist-two operators. It is valid for small, but still finite, coupling constant, as long as the anomalous dimensions of the twist-two operators are small enough that higher order terms in the expansion (\ref{A1CBE}), from higher twist operators, are subleading.
 Thus, in this equation at finite coupling, terms with different $J$ can not mix because they have different powers of $u$. 
This means that
\begin{align}
&12\left(\frac{1}{z^2}-\frac{1}{z}+\frac{1}{6}\right)  a_J\,   F\!\left(1+J+\frac{\gamma}{2},z\right)=
\label{RelationOPEnosum}
\\
&  b_{J+2} \,\frac{  z^2}{4}\, F\!\left(3+J+\frac{\gamma}{2},z\right)
+b_J'\, F\!\left(1+J+\frac{\gamma}{2},z\right)+
b_{J-2}'' \, \frac{4}{z^2}\, F\!\left(J-1+\frac{\gamma}{2},z\right) ,
\nonumber
\end{align} 
for all $J=2,4,6,\dots$. Analyzing the Taylor expansion in $z$ on both sides, we see that this equation is satisfied if 
\begin{align}
&b_{J-2}''=3a_J \,, \ \ \ \  \ \ \ \ \  
 b_{J}'=\frac{\big(\gamma(J) +2J\big) \big(\gamma(J)+2 J+2\big)}{2\big(\gamma(J) +2 J-1\big) \big(\gamma(J) +2 J+3\big)}\,a_J \,,
 \nonumber\\
&b_{J+2} =\frac{3\big(\gamma(J) +2 J+2\big)^2 \big(\gamma(J) +2 J+4\big)^2}{16 \big(\gamma(J) +2 J+1\big) \big(\gamma(J) +2 J+3\big)^2 \big(\gamma(J) +2 J+5\big)}\,a_J\,.
   \label{eq:OPErelation}
\end{align}
These relations are non-perturbative. They follow from supersymmetry as explained in \cite{Dolan:2001tt}.  
Notice that for $J=0$, we find $b_2=4 a_0/15$ which is compatible with (\ref{a0b2}).

\subsection{Higher loop prediction for leading logs}
\label{OPElogs}

The conformal block expansion (\ref{CBexpansion}) can also be used to extract the 
leading behaviour  of the four-point function in the $u\rightarrow 0$  limit.
More precisely, comparing equations  (\ref{A20WCE}) and  (\ref{A20CBE}) it is easy to conclude that the leading log at any loop order is given by  
\begin{align}
\theta^{(l)}_0(1-z) 
 &= \frac{  4^l (1-z) }{l!\, z^2}
\sum_{J=0 \atop {\rm even} }^\infty
\frac{J!^2}{(2J)!} \, \big(S_1(J) \big)^l
z^J  F ( 2+2J,z) \,, 
  \label{LLOGlorOPE}
\end{align}
which only involves tree-level OPE coefficients and one-loop anomalous dimensions.
In fact, at $n$-loop order, the anomalous dimension
and OPE coefficients at the same $n$-loop order only enter through the linear and constant
terms in the $\log u$ expansion. All the other higher $\log u$ terms are determined by 
lower order anomalous dimension and OPE coefficients.
For example, 
one can use information on anomalous dimensions up to four loops \cite{DressingWrapping,4loopsTwist2}, and on OPE coefficients 
up to three loops \cite{Eden:2012rr}, to fix the small $u$ behaviour up to the $(\log u)$-independent term $\theta_4^{(4)}$.
In appendix \ref{app:IntRep}, we give an integral representation for the sum in (\ref{LLOGlorOPE}).

Using formula (\ref{LLOGlorOPE}) we can predict the leading log behaviour at four loops,
\begin{align}
\tilde{\k}_{0}^{(4)}&=\frac{8}{6} \left(3H_{1,3}+3H_{2,2}+ H_{3,1}+3H_{1,1,2}+2H_{1,2,1}+2
   H_{2,1,1}+3
   H_{1,1,1,1}\right) , \\
{\kappa}_{0}^{(4)}&=\frac{8}{3} \left( H_{4}- H_{1,3}- H_{2,2}\right),
\end{align}
where we used the notation (\ref{kappa}) to write the result.
Using the known form of the anomalous dimensions of the leading twist operators up to four loops and OPE coefficients up to three loops, we can also predict  the sub-leading logs $\theta_k^{(4)}$ for $k=1,2,3$. We give these predictions in appendix \ref{ap:4loopsOPE}. 

We shall see in the next section  that in the Regge limit of the four-point function a similar strategy can be used to fix some leading logs 
from knowledge of the BFKL spin and Regge residue.

\section{
Regge limit}
\label{sec:ReggeLimit}

\begin{figure}[t!]
\begin{centering}
\includegraphics[scale=0.6]{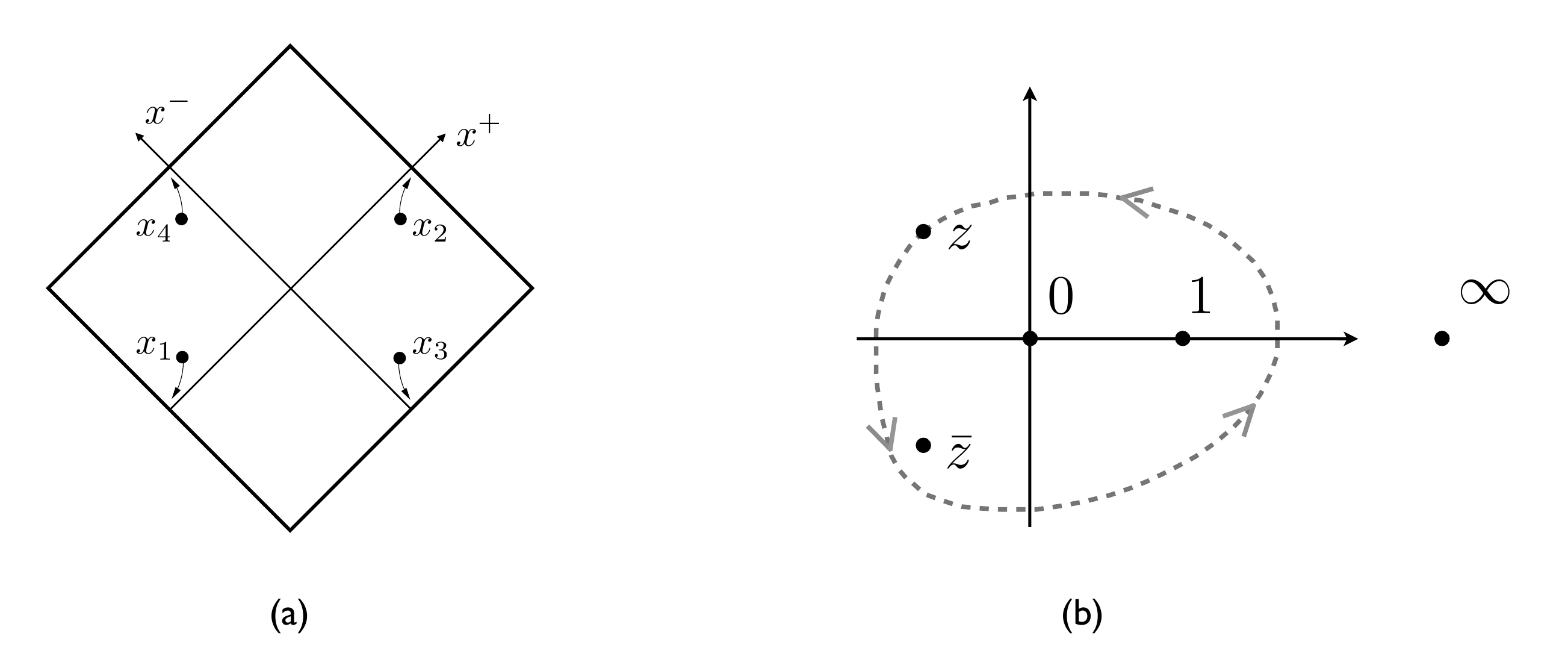}
\par\end{centering}
\caption{\label{figReggeLim} (a) 
Regge limit shown in a conformal compactification of the light cone plane. The positions of the operators $x_i$ go to null infinity as shown.
(b) The analytic continuation in $z$ and $\bar{z}$, starting from the Euclidean correlation function.}
\end{figure}


In a conformal field theory the Regge limit of a four-point function is obtained from a specific Lorentzian kinematical limit
where all the points are taken to null infinity \cite{CornalbaRegge,ourBFKL}. This limit can be defined by 
$x_1^+ \to \lambda x_1^+ $, 
$x_2^+ \to \lambda x_2^+ $, 
$x_3^- \to \lambda x_3^- $, 
$x_4^- \to \lambda x_4^- $ and $\lambda \to \infty$, 
keeping the causal relations $x_{14}^2, x_{23}^2 <0$. We will choose  all the other $x_{ij}^2>0$,  as show in figure \ref{figReggeLim}, although this is not essential.
This Lorentzian regime, needed to take the  Regge limit, can be obtained from the Euclidean regime by analytic continuation.
This is done by fixing $\bar{z}$ and by analytically continuing in $z$ counter clockwise around $0$ and $1$, as also
show in figure \ref{figReggeLim}. The analytically continued function defines the Lorentzian correlation function. Physical
space-time points now correspond to both $z$ and $\bar{z}$ real numbers. For the specific form we took the Regge limit,
which kept $x_{14}^2, x_{23}^2 <0$, we have both $z$ and $\bar{z}$ small and positive. It is then 
convenient to introduce the  variables $\sigma$ and $\rho$ via
\be
z = \sigma e^\rho\,, \qquad \bar z = \sigma e^{-\rho}\,.
\ee
They are related to the  cross ratios $u$ and $v$ defined in (\ref{CrossRatios}) by
\be
u=\sigma^2\,,\ \ \ \ \ \ \ 
v=(1-\sigma e^\rho)(1-\sigma e^{-\rho})\approx 1-2\sigma \cosh \rho\,.
\ee
The  Regge limit corresponds now to $\sigma \to 0$ with fixed $\rho$.

\subsection{Weak coupling expansion}
\label{Reggeweakcoupling}

In the Regge limit the  general structure of the function $F(u,v)$,  with loop expansion defined in (\ref{ExpansionF}), is given by  
\be
u  F^{(l)}(u,v)=   \sum_{k=0}^{l-2} 
(\log \s)^{l-2-k} \x_{k}^{(l)}(\r) + O(\s)\,,
 \label{weakRegge}
\ee
where $k=0$ is called leading log, $k=1$ next-to-leading log, etc. The invariance of $F(u,v)$ under the interchange of $z$ and $\bar{z}$ implies that   $\x_{k}^{(l)}(\r)=\x_{k}^{(l)}(-\r)$.
Note that  only the two-loop and higher contributions to $F(u,v)$ enter in  (\ref{weakRegge}). In fact, 
if we take the Regge limit on the form of $F(u,v)$ up to three loops (\ref{Fuvto3loops}), we see that only the functions $g$ and $h$ contribute,
\begin{align}
u  F(u,v) = 
 \frac{r}{(1-r)^2}  \Bigl( g(z,\bar z)  + g\bigl(\tfrac{1}{1-z},\tfrac{1}{1-\bar z} \bigr) \Bigr) - \frac{r}{1 - r^2}  \,h(1-z, 1-\bar z) +O(\sigma) \,,
\label{Reggelimform}
\end{align}
where $r=z/\bar{z} = e^{2 \rho}$.
Since $g$ and $h$ vanish at one loop, the four-point function at this order has a subdominant behavior in the Regge limit $\s \to 0$. 

The two-loop contribution was calculated in \cite{ourBFKL}, where it was found that
\be
u  F^{(2)}(u,v) =
\xi^{(2)}_0(\rho)+ O(\sigma) 
= - \frac{8 r}{(1-r)^2}  \,\pi^2 \log^2 r + O(\sigma)\,.
\label{xi00}
\ee
As expected, the result is invariant under $r \to 1/r$.

At three loops one finds from the explicit results of \cite{Drummond:2013nda} that the coefficient of the log-divergent term is given by\footnote{More details on the analytic continuation are given in Appendix \ref{AnalyticContinuation}.}
\be
\xi^{(3)}_0(\rho) = -\frac{32r}{(1-r)^2}\, \pi^2 \big(H_{0,0,0} + 2 H_{1,0,0}-2\zeta_3\big)\,,
\label{xi10}
\ee
where the argument of the harmonic polylogarithms is $r=e^{2 \rho}$ and  $\zeta_x=S_{x}(\infty)$ is   the Riemann zeta function. The coefficient of the finite term is given by
\begin{align}
 \xi^{(3)}_1(\rho) =&\   64 \pi ^2\frac{r}{\left(1-r\right)^2}\bigg(
H_{0,0,0,0}(r)+2 H_{1,0,0,0}(r)+4
   H_{1,1,0,0}(r)+2 H_{2,0,0}(r)
    \nonumber\\
&
-2 \zeta_3\left( H_0(r)+2 H_1(r)\right) +
     \frac{\zeta_2}{2} H_{0,0}(r)-  3\zeta_4
    \bigg) + \frac{i\pi}{2}\x_0^{(3)} .
\label{xi11}
\end{align}

Note that, although at three loops the function $h$ is non-zero, in the Regge limit its analytic continuation is power suppressed as $\sigma \rightarrow 0$. Therefore at three loops we do not see any contribution to the four-point function with the form of the second term in (\ref{Reggelimform}), with its prefactor $r/(1-r^2)$. We will see, from the analysis we present in sections \ref{Pomeron} - \ref{NLL4loop}, that at four loops there is such a contribution.

\subsection{Pomeron spin and residue}
\label{Pomeron}

The Regge limit of CFT four-point functions   has been studied in \cite{CornalbaRegge,ourSW, ourCPW, ourEikonal,ourBFKL,ourDIS,Costa:2012cb}.
By studying the Regge limit of the conformal block expansion, the following general form was derived
\cite{CornalbaRegge, Costa:2012cb}
\begin{align}
A_1 
\approx 
2\pi i \sum_t
\int\! d\nu \,\alpha_t\!\left(\nu\right)\sigma^{1-j_t\left(\nu\right)}\,\Omega_{i\nu}(\rho)\,.
\label{eq:alphat}
\end{align}  
where $t$ labels Regge trajectories and this is valid in the Regge limit $\s \to 0$.
The function $\Omega_{i\nu}$ in (\ref{eq:alphat}) is a harmonic function in three-dimensional hyperbolic space and it is explicitly given by
\begin{align}
\Omega_{i\nu}(\rho)=\frac{\nu\sin(\rho\nu)}{4\pi^2\sinh\rho}\,.
\label{eq:Omega}
\end{align}
We shall refer to the functions $j(\nu)$ and $\alpha(\nu)$ as the Reggeon spin and residue, respectively.

Generically, there is a leading trajectory, with maximal intercept $j_t(0)$, and the other trajectories describe subleading corrections at small $\s$.
This is believed to be the case in planar SYM at finite 't Hooft coupling, where the leading Regge trajectory is called the pomeron.
However, at weak coupling we expect many trajectories to have $j(\nu)=1+O(g^2)$, of which the pomeron is the simplest example.\footnote{See for instance \cite{Korchemsky:2003rc}.  We thank Benjamin Basso for calling our attention to this accumulation of Regge trajectories in the weak coupling limit.}
The main difference in these trajectories is that $\a(\nu)= O(g^{2n})$ for a $n-$Reggeon trajectory ($n=2$ for the pomeron), and only even $n$ contributes to the singlet channel $A_1$.
Therefore, subleading Regge trajectories only contribute to subleading logs at four-loops and higher. In the notation of (\ref{weakRegge}), they only affect $\x_k^{(l)}$ for $l \ge 4$ and $k \ge 2$.
Keeping in mind this subtlety, from now on, we shall drop the label $t$ and write 
\begin{align}
A_1 
\approx 
2\pi i  
\int\! d\nu \,\alpha \!\left(\nu\right)\sigma^{1-j \left(\nu\right)}\,\Omega_{i\nu}(\rho)\,.
\label{eq:alphaPom}
\end{align}
keeping only the pomeron trajectory.
This is sufficient to describe leading and subleading logs at any loop order.

We write the expansion of the pomeron spin $j(\nu)$ and the residue $\alpha(\nu)$ in the usual form
\be
j(\nu)= 1+\sum_{n=1}g^{2n} j_n(\nu)\,,\ \ \ \ \ \ 
\alpha(\nu)=\sum_{n=2}g^{2n}\alpha_n(\nu)\,.
\ee
Comparing equations (\ref{eq:alphaPom}), (\ref{weakRegge}) and expression (\ref{A1A20}) for $A_1$, we conclude that
\begin{align}
\frac{8}{N^2-1} \,\x^{(2)}_0(\r)&=
2\pi i  
\int\! d\nu \,\alpha_2 (\nu)\,
\Omega_{i\nu}(\rho) \,, 
\label{alpha2}
\\
\frac{8}{N^2-1}\, \x^{(3)}_0(\r)&=
-2\pi i  
\int\! d\nu \,\alpha_2 (\nu)\,
j_1(\nu)
\,\Omega_{i\nu}(\rho)\,,
\\
\frac{8}{N^2-1} \,\x^{(3)}_1(\r)&=
2\pi i  
\int\! d\nu \,\alpha_3(\nu)
\,\Omega_{i\nu}(\rho)\,.
\label{alpha3}
\end{align}
Using the explicit expressions (\ref{xi00}), (\ref{xi10}) and (\ref{xi11}) for the Regge limit of the four-point function up to three loops, we can determine the functions  $\a_2(\nu)$,  $\a_3(\nu)$ and $j_1(\nu)$ by inverting the integral transform above as explained in \cite{CornalbaRegge,Mythesis}.
In fact, the pomeron spin is known at leading order  and next-to-leading 
order\footnote{Notice that the BFKL spin at NLO starts to contribute to the four-point function of the stress-energy tensor multiplet only at four loops.} 
for quite a long time  
\cite{Kotikov:2000pm,KotikovLipatov02}
\begin{align}
j_1(\nu)& =  8\Psi(1)- 4\Psi\!\left(\frac{1+i\nu}{2}\right)-4 \Psi\!\left(\frac{1-i\nu}{2}\right) ,
\label{BFKLspinj}
\\
j_2(\nu)&= 4j_1''(\nu)+24\zeta_3 - 2\zeta_2 j_1(\nu)\,
-8\Phi\!\left(\frac{1+i\nu}{2}\right) -  8\Phi\!\left(\frac{1-i\nu}{2}\right),
\label{BFKLspinj2}
\end{align}
where $\Psi(x)=\Gamma'(x)/\Gamma(x)$ is the Euler $\Psi$-function and  
\be
\Phi(x)=\sum_{k=0}^{\infty} \frac{2}{k+x}\sum_{r=0}^{\infty}\frac{(-1)^{r+1}}{(k+1+r)^2} \,.
\ee
The function $\alpha(\nu)$ was computed at leading order in \cite{ourBFKL} and it is given by
\begin{equation}
\alpha_2(\nu)=i\,
\frac{16\pi^{5}  }{N^{2} -1}
\frac{  \tanh\!\left(\frac{\pi \nu}{2} \right)}
{ \nu\cosh^{2}\!\left(\frac{\pi\nu}{2}\right)}
\,.
\label{LOalpha}
\end{equation}

Next we consider the computation of the next-to-leading order pomeron residue $\alpha_3(\nu)$. 
Up to next-to-leading order (at least), the functions $j(\nu)$ and $\a(\n)$ obey the principle of maximal transcendentality. To see that more explicitly we write
\be
j(\nu)=1+\sum_{n=1}^\infty g^{2n} \left[F_n\!\left(\frac{i\nu-1}{2}\right)+F_n\!\left(\frac{-i\nu-1}{2}\right)\right] ,
\label{eq:jstruc}
\ee
and
\be
\alpha\!\left(\nu\right)=\frac{\pi^2}{N^{2} -1}
\frac{1}{\nu}\sum_{n=2}^\infty g^{2n} \left[G_n\!\left(\frac{i\nu-1}{2}\right)-G_n\!\left(\frac{-i\nu-1}{2}\right)\right] ,
\label{eq:alphastruc}
\ee
so that $F_n$ and $G_n$ have transcendentality $2n-1$ and can be written in terms of harmonic sums. In particular, for
the above pomeron spin formulae (\ref{BFKLspinj}) and (\ref{BFKLspinj2}), and for the pomeron residues determined from integrals (\ref{alpha2}) and (\ref{alpha3}), 
we have
\begin{align}
F_1(x)=&\, 
-4S_1 \,,\\
F_2(x)=&\ 
4\left( \pi^2 \ln2-\frac{3}{2}\zeta_3 
+\frac{\pi^2}{3} S_1 +\pi^2 S_{-1} +2S_3 
-4S_{-2,1}  \right)  , \\
G_2 (x )=&\ 
16 S_3 \,,\\
G_3(x) =& \, 
-128\left(2 \zeta_3   S_2-4\zeta_3 S_{1,1}-2 S_{1,4}-2
   S_{2,3}+4
   S_{1,1,3}+3\zeta_4\, S_1+S_5+\frac{\zeta_2}{2}\,S_3
    \right), \label{eq:G3}
   \\ &+ i64\pi \big(S_4+2\zeta_3S_1-2S_{1,3}\big)\,,
   \nonumber
\end{align}
where all harmonic sums have argument $x$.

In \cite{Balitsky:2009yp}  the next-to-leading order pomeron residue $\alpha_3(\nu)$ was computed using the techniques of operator expansion over colour dipoles \cite{Balitsky:1995ub}. 
However, the proposed expression  is
different from our result obtained from taking directly the Regge limit of the three-loop contribution to the four-point function.
We are confident that our result is correct, in part, because of the non-trivial consistency check that we will perform in section \ref{sec:CRT}.
Nonetheless, we are not able to pinpoint any specific mistake in the calculations of 
\cite{Balitsky:2009yp}. It would be interesting to return to this question using the recent works \cite{Caron-Huot:2013fea,Balitsky:2013npa}.

Finally, let us just make a technical remark, that will be useful  when analysing the residue of $\alpha_3(\nu)$ near $J=1$.
It turns out that, using expression (\ref{eq:newsymmetryrel}) given in  appendix \ref{App:Harmonics},  
we can express $\alpha_3(\nu)$ in terms of simpler functions
\begin{align}
\alpha_3(\nu)&=\frac{1}{4}  \alpha _2 \left(j_1^2+16\zeta_2\right)-16 i \pi ^7\,
\frac{\sinh \!\left(\frac{3  \pi  \nu }{2}\right)-11 \sinh\! \left(\frac{\pi   \nu }{2}\right)}{3\nu\cosh^5\!\left(\frac{\pi  \nu
   }{2}\right)}  +\frac{i \pi}{2}j_1\alpha_2\,.
\label{eq:alpha3simple}
\end{align}
Notice that the first two terms are imaginary and the last term is real because $\alpha_2(\nu)$ is imaginary for real $\nu$.

\subsection{Higher-loop prediction for leading logs}
\label{LL}

In section \ref{OPElogs} we saw that for  the light-like OPE limit the leading log dependence in $u$ at any loop order
is fixed by the LO anomalous dimensions and OPE coefficients. Something analogous occurs in the Regge limit. In fact,
the general form of the amplitude  (\ref{eq:alphaPom}) implies that knowledge of the 
leading order spin $j_1(\nu)$ and residue $\alpha_2(\nu)$, automatically fixes the 
leading $\log \sigma$ dependence of the amplitude at any loop order. 
More precisely, for $n\ge2$ we have
\begin{align} 
\frac{8}{N^2-1} \, \x^{(n)}_0(\r)&= \frac{2\pi i  }{(n-2)!}
\int\! d\nu \,\alpha_2 \!\left(\nu\right)
\left[-j_1(\nu)\right]^{n-2}
\,\Omega_{i\nu}(\rho) \,.
\end{align}
Using expressions (\ref{BFKLspinj}), (\ref{LOalpha}) and (\ref{eq:Omega}), we can write
\be
\x^{(n)}_0(\r)= -\frac{\pi^4 4^{n-2}}{(n-2)!\sinh \r} \,I_{n-2}(\r) \,,
\ee
where
\begin{align}
I_n(\r) &=\int d\nu \,
 \frac{\tanh\!\left(\frac{\pi\n}{2}\right)}
 {\cosh^2\!\left(\frac{\pi\n}{2}\right)} 
 \left[S_1\left(\frac{i\n-1}{2}\right)
 +S_1\left(\frac{-i\n-1}{2}\right)
 \right]^{n}
 \sin(\nu\rho) \,.
\end{align}
In appendix \ref{App:LLOGREGGE}, we show that this integral obeys a recursion relation. The final result reads
\be
\x^{(n)}_0(\r)= -\frac{\pi^2 4^{n-1}}{(n-2)!} \frac{r}{1-r} \,B_{n-2}(r)\,,
\ee
where  $r=e^{-2\r}$ and  $B_n$ obeys the recursion relation  
\be
B_n(r) = \int_0^1 \frac{ dx }{1-x } \left[
2B_{n-1}(r)-  B_{n-1}\left(rx\right)
-  \frac{1}{x}B_{n-1}\left(\frac{r}{ x}\right)
\right] ,
\ee
with initial condition
\be
B_0(r)= \frac{\log^2r}{1-r}\,.
\ee
The solution of this recursion relation can always be written explicitly in terms of harmonic polylogarithms.
For example, at four and five loops we predict that
\begin{align}
\x^{(4)}_0(\r)&=\frac{64\pi^2r}{(1-r)^2}\big( H_{0,0,0,0} + 4 H_{2,0,0} + 4
   H_{1,0,0,0} + 8 H_{1,1,0,0}   - 4 \zeta_3
   (H_0+2H_1)-6\zeta_4\big) \,,\\
\x^{(5)}_0(\r)&=\frac{256 \pi^2 r}{3(1 - r)^2}\Big( 
  H_{0,0,0,0,0} +6(4H_{1,2,0,0} +2 H_{2,0,0,0} +4 H_{2,1,0,0}  +H_{1,0,0,0,0} + 
         4 H_{1,1,0,0,0})\nonumber\\
&+8(H_{3,0,0}+6 H_{1,1,1,0,0}) - 6\zeta_3(4H_{2} + H_{0,0}+ 4H_{1,0}+ 8 H_{1,1}) - 6\zeta_4 (H_{0} + 6H_{1})-6\zeta_5\Big)\,,
\nonumber
\end{align} 
where we omitted the dependence of the  harmonic polylogarithms in the variable $r$.

In general, we expect the $\x^{(n)}_0$ to be given by the product of the rational pre-factor $\pi^2 r/(1-r)^2$ times a linear combination of harmonic polylogarithms of transcendentality $n$ and invariant under $r \to 1/r$.

\subsection{Next-to-leading log at four loops}
\label{NLL4loop}

Since we know the NLO Regge spin and residue, we can also express the next-to-leading log (NLL) behaviour in $\log\sigma$ as a transform with respect to the harmonic function $\Omega_{i\nu}$. Let us consider for simplicity
 the case of four loops. In this case the NLL term is given by the following integral 
\begin{align}
\frac{8}{N^2-1} \, \x^{(4)}_1(\r)=-2\pi i
\int\! d\nu\Big(j_2(\nu)\alpha_2(\nu)+j_1(\nu)\alpha_3(\nu)\Big) \Omega_{i\nu}(\rho)\,.
\label{eq:NLLprediction} 
\end{align}
The integral over $\nu$ can be done using integral representations for the BFKL spin and residue, or just by deforming the contour picking poles and matching to an appropriate ansatz, 
with result 
\begin{align}
\x_1^{(4)}&=\frac{64\pi^2 r}{\left(r^2-1\right)}\left(2H_{3,0,0}+ 2H_{2,0,0,0}+ 4 H_{2,1,0,0}+\zeta_2 H_{0,0,0} -2\zeta_3(2H_{2} + H_{0,0})+\frac{5\zeta_4}{4}H_{0} - 2\zeta_5\right)\nonumber\\
&+
\frac{64\pi^2 r}{\left(1-r\right)^2}
\Big(
   H_{0,0,0,0,0}+2H_{3,0,0}+\zeta_2 (H_{0,0,0}+2H_{1,0,0})+\zeta_3\left(H_{0,0}-2\zeta_2\right)-6\zeta_4 H_{0}-12 \zeta_5\Big)\,\nonumber\\
&+\frac{256\pi^2 r}{(1-r)^2}\Big(H_{0,0,0,0,0}+ 4H_{3,0,0}+4H_{1,0,0,0,0} +6(2 H_{1,2,0,0} +H_{2,0,0,0}-2H_{2,1,0,0}+4H_{1,1,1,0,0}  \nonumber\\
&+\big.2H_{1,1,0,0,0})+\frac{\zeta_2( H_{0,0,0}+2H_{1,0,0}+2\zeta_3)}{2}-4\zeta_3(3H_{2}+H_{0,0}+ 3H_{1,0}+6H_{1,1})
\nonumber\\
&-3\zeta_4( H_{0}+6H_{1})-4\zeta_5 \Big)+\pi i\x^{(4)}_0\,,
\end{align} 
where the first two lines correspond to the first term in (\ref{eq:NLLprediction}) and the last three to the remaining part. 
Notice that there is a different prefactor which comes precisely from the $\Phi$ function in the Regge spin $j_2(\nu)$ in (\ref{BFKLspinj}), 
that can be written using Harmonic Sums with negative indices. The term proportional to $\x_0^{(4)}$ in the last line is imaginary.

\section{Conformal Regge Theory}
\label{sec:CRT}

In the previous two sections, we considered two kinematical limits of the four point function of the stress tensor multiplet in SYM. 
A priori, these two limits seem unrelated. However, this is not the case.
The connection between the two limits can be established through Conformal Regge Theory \cite{Costa:2012cb}.\footnote{The existence of a relation was known before  \cite{KotikovLipatov02}, but not the precise form of equation (\ref{eq:alphaOPE}) given below.}
We shall not repeat the complete argument made in 
\cite{Costa:2012cb} but we will briefly review the main steps involved.

The starting point is the 
conformal block decomposition (\ref{CBexpansion}) written as 
\be
A_R(u,v) =
\sum_{J=0}^\infty \int_{-\infty}^\infty  d\nu   \,
\frac{i \nu \, d_J^R(\nu^2)}{\pi K_{2+i\nu,J}}
\, G_{2+i\nu,J}(u,v)\,,
\label{CBexpansion2}
\ee
where 
\be
K_{\Delta,J}=\frac{\Gamma(\Delta+J)\,\Gamma(\Delta-1+J)  }{4^{J-1}\Gamma^6\!\left(\frac{\Delta+J}{2}\right)\Gamma^2\!\left(\frac{4-\Delta+J}{2}\right)
}\,.
\ee
To reproduce the exchange of an operator of dimension $\Delta$ and spin $J$, as written in  (\ref{CBexpansion}),  
the partial amplitude must have the singular behaviour
\be
d_J^R(\nu^2)\approx a_{\Delta,J}^R\,\frac{ K_{\D,J}}{\nu^2+(\Delta-2)^2}\,,
\label{polesfromCB}
\ee
such that we recover (\ref{CBexpansion}) by closing the integration contour in the lower half of the $\nu$ complex plane and perform the integral by residues.

From now on, we restrict our attention to the singlet channel $A_1$.
The next step is to consider the 
Sommerfeld-Watson transform, keeping only the operators in the leading Regge trajectory, for which we have a relation $\Delta=\Delta(J)=2+J+\g(J-2)$ in the notation of section \ref{sec:AnomDim}.
Analytically continuing in the spin $J$, the Sommerfeld-Watson integration over $J$ can be done by picking the Regge pole $j(\nu)$ defined by 
\begin{align}
\nu^2+\left(\Delta\big(j(\nu)\big)-2\right)^2=0\,,
\label{eq:BFKLInverD}
\end{align} 
and we obtain the final result (\ref{eq:alphaPom}) 
where
\be
\alpha(\nu)=-\frac{\pi^2 2^{j(\nu)-3}e^{i\pi j(\nu)/2}}{\nu\sin\left(\frac{j\left(\nu\right)}{2}\right)}\,
\gamma(\nu)\gamma(-\nu)j'(\nu)K_{2+ i\nu,j(\nu)} \,b_{ j(\nu)}\,,
\label{eq:alphaOPE}
\ee
with $b_J$ defined in (\ref{A1CBE}) and
\be
\gamma(\nu)=\Gamma^2\!\left(\frac{2+j(\nu)+i\nu}{2}\right)\,.
\ee
We  refer to the function $\alpha(\nu)$ as the Regge residue, since it is related to the residue of the dominant Regge pole  
that follows from (\ref{polesfromCB}).

Equations (\ref{eq:BFKLInverD}) and (\ref{eq:alphaOPE}) establish a precise relation between the two kinematical limits studies in the previous sections. The pomeron spin $j(\nu)$ and residue $\alpha(\nu)$ are related to analytic continuations of the dimensions $\D(J)$ and (square of) OPE coefficients $b_J$ of the leading twist operators. 
Since relation (\ref{eq:BFKLInverD}) does not commute with perturbation theory it is possible to obtain all-loop predictions for the function  $\Delta(J)$ 
from the BFKL spin $j(\nu)$, and vice versa. This was explored at weak coupling in \cite{DressingWrapping} and at strong coupling in \cite{Costa:2012cb}. 
 In fact (\ref{eq:BFKLInverD}) was crucial in determining perturbative information about the anomalous dimension at four loops. 
 The mismatch between the prediction coming from (\ref{eq:BFKLInverD}) and $j\!\left(\nu\right)$, and known data at four loops computed from 
 the asymptotic Bethe ansatz, was resolved by including wrapping effects \cite{4loopsTwist2}. 
 Thus, both eqs. (\ref{eq:BFKLInverD})  and (\ref{eq:alphaOPE})  can be used as a consistency check on the available data in both regimes.
In particular, they also provide a new test that direct computations of OPE coefficients must pass.

\subsection{Prediction for OPE coefficients}

In this section we shall use equation (\ref{eq:alphaOPE}) and the perturbative expressions of $\a(\nu)$ given in section \ref{sec:ReggeLimit} to derive all loop predictions for the square of OPE coefficients, $b_J$.
As explained in \cite{Costa:2012cb}, we take in  (\ref{eq:alphaOPE})  the double limit $g^2\rightarrow 0$ and $J\rightarrow 1$, with fixed  ratio $g^2/(J-1)$,  
and write the square of the OPE coefficients in an expansion around $J=1$ as
\be
b_J=\left(J-1\right)f\!\left(\frac{g^2}{J-1}\right)+
\left(J-1\right)^2h\!\left(\frac{g^2}{J-1}\right)+O\!\left((J-1)^3\right),
\label{eq:OPEexpansion}
\ee
with
\be
f(x)=\sum_{n=0} f_n\,x^n \,,\ \ \ \ \ \ \ \
h(x)=\sum_{n=0} h_n\,x^n\,.
\ee
Then, expanding  (\ref{eq:alphaOPE}) near $\nu=-i$, we obtain a 
 system of equations for the coefficients $f_n$ and $h_n$. With the knowledge of $\a_2(\nu)$ and $\a_3(\nu)$, this can be solved for all $f_n$ and $h_n$, with the  first terms given by 
 \begin{align}
&f_0=\frac{2}{3}\,, \ \ \ \ \ f_1=\frac{64}{9}\,, \ \ \ \ f_2=\frac{32}{27}\left(61-3\pi^2\right), \ \ \ \ f_3=\frac{256}{81}\left(223-12\pi^2-27\zeta_3\right),
 \nonumber\\
&h_0=\frac{2}{9}(-8+3\ln2)\,, \ \ \ \ \ \ \ \ \ \ \ \ h_1=\frac{4}{27}(-244+9\pi^2 +48 \ln 2)\,, \nonumber\\
&h_2=\frac{16}{27} \left(153 \zeta_3-892+122 \ln2-2 \pi ^2 (3\ln2-20)\right) ,\\
&h_3=\frac{64}{1215}\,\Big(20 \big(669 \ln2-27 \zeta_3 (3\ln2 -44)-6320\big)+171 \pi ^4+\pi ^2 (6405-720 \ln2)\Big)
\nonumber\,.
\end{align}
Equations (\ref{eq:OPErelation}) and (\ref{eq:aJ}), together with the explicit results for the OPE coefficients up to 
three loops derived in \cite{Eden:2012rr} and
reviewed in appendix \ref{App:GammaOPEcoefs}, can be used to  check that all these coefficients are indeed correct. 
This check is extremely non-trivial, therefore confirming our NLO computation of the Regge residue $\alpha_3(\nu)$ in (\ref{eq:G3}).
Moreover, forthcoming computations of OPE coefficients at higher loops  must  pass this 
test of conformal Regge theory. We state here the prediction for four loops
\begin{align}
f_4=&\ \frac{512 (15800-915 \pi ^2-36 \pi ^4)}{1215}\,,
\nonumber\\
h_4=&\ -\frac{256\big(9 \pi ^4 (24 \ln2-221)+90 \pi ^2 (81\zeta_3-669+61 \ln 2)\big)}{3645}\\
&\ +\frac{1280 (21870 \zeta_5-218720+18960 \ln 2+31293 \zeta_3)}{3645}\,.
\nonumber
\end{align}

\section{Conclusion}
\label{sec:Conclusion}

We have studied the Regge limit and the Lorentzian OPE limit of the four-point function of stress tensor multiplets in SYM up to three loops. Its consistency with the OPE had previously been verified in \cite{Eden:2012rr} by taking limits on the individual integrals. Here we have verified its consistency with conformal Regge theory \cite{Costa:2012cb} by performing the relevant analytic continuation of the integrated result. In particular we were able to verify the leading log behaviour predicted by \cite{Costa:2012cb} and derive the next-to-leading log behaviour from the known analytic form of the four-point function at three loops \cite{Drummond:2013nda}. Both the leading and next-to-leading log contributions respect maximal transcendentality.   With the leading and next-to-leading log contributions in the Regge limit to hand we were able to make predictions at all orders in the 't Hooft coupling for the leading and next-to-leading contributions to the three-point functions of two stress tensors and one twist-two operator in the limit $J\rightarrow 1$.

Using our results for the Regge limit of the four-point function at three loops, we were also able to predict the behaviour of the four-point function in both kinematical limits at higher loop orders. These predictions place strong constraints on the form of the four-point function at higher loops. We hope that they can be used as input for perturbative bootstrap of the kind that has been successfully applied to scattering amplitudes in planar $\mathcal{N}=4$ super Yang-Mills theory in \cite{Dixon:2011pw,Dixon:2011nj,Dixon:2013eka}. Indeed, the problems of bootstrapping the correlation function and scattering amplitudes are quite analogous. In both problems we have analytic constraints from an operator product expansion; in the case of the correlator this is just the Lorentzian OPE we studied in Section \ref{sec:lorOPE}, while for the scattering amplitude it is the OPE of the dual light-like Wilson loop \cite{Alday:2010ku}. Also for both problems there is a Regge limit which, at least perturbatively, provides different information from the OPE.

In order to investigate the constraints imposed by the Lorentzian OPE and Regge limits on the correlation function at higher loops one must first construct an ansatz. The ansatz consists of two parts; firstly one needs the leading singularities, then one needs to specify the relevant class of pure transcendental functions. In the case of the three-loop correlator, the leading singularities are the rational functions of $z$ and $\bar z$ appearing as prefactors of the functions $f$, $g$ and $h$ in (\ref{Fuvto3loops}), while the pure transcendental functions are $f$, $g$ and $h$ themselves. 

It is tempting to conjecture that no new leading singularities appear beyond three loops, i.e. that the four-point function can be written in the form (\ref{Fuvto3loops}) to all orders, with the functions $f^{(l)}$, $g^{(l)}$ and $h^{(l)}$ given by linear combinations of single-valued multiple polylogarithms of weight $2l$. The reason is that it is not obvious that one can introduce new prefactors, algebraic functions of $z$ and $\bar z$, which respect crossing symmetry and reduce to the required form in the Lorentzian OPE limit. However we have not investigated this question in detail and it could be that leading singularities of a different form do arise at some loop order. Ultimately this question can be answered by studying the form of the integrand at higher loops found in \cite{Eden:2012tu}.

The relevant class of transcendental functions was partly uncovered in \cite{Drummond:2013nda}. In addition to the letters $\{z,\bar z,1-z,1-\bar z\}$ appearing in the symbol of the four-point function up to two loops, a fifth letter $z-\bar z$ appears at three loops. Furthermore, a specific four-loop integral was evaluated using techniques developed in \cite{Drummond:2012bg}, and another letter $1-z\bar z$ was needed. Further letters related by crossing symmetry are therefore needed to describe the same integral in different orientations. Thus a minimal ansatz for the class of transcendental functions is that they are given by single-valued multiple polylogarithms whose symbols are described by the set of eight letters $\{z,\bar z,1-z,1-\bar z,z-\bar z, 1-z\bar z,1-z-\bar z,z+\bar z-z\bar z\}$.

It would be interesting to know whether the above ansatz is sufficient to admit a solution compatible with the OPE and Regge limit and if so, how many undetermined coefficients remain after imposing the constraints. We leave this exercise for the future.

Another interesting direction for future investigation is to consider four-point functions of $\tfrac{1}{2}$-BPS operators of the higher $SU(4)$ representations $[0,p,0]$. Up to two loops such correlators are given in terms of the same integrals appearing in the stress-energy correlation function \cite{Arutyunov:2003ad}. It would be interesting to see if a similar pattern continues at higher loops and whether such questions can be answered by applying consistency of OPE and Regge limits against an ansatz.

\section*{Acknowledgements}

We would like to thank Benjamin Basso, Pedro Vieira and Evgeny Sobko for discussions.
J.P. and V.G.  wish to thank CERN for the great hospitality during our visit in the spring of 2013 when   this work was initiated.
M.S.C and V.G also thank IPMU at Tokyo University for the great hospitality during the progress of this work. J.P. is grateful to Perimeter Institute for the hospitality in July of 2013.
The research leading to these results has received funding from the [European Union] Seventh Framework Programme [FP7-People-2010-IRSES] under grant agreements No 269217,
317089.
This work was partially funded by the grant CERN/FP/123599/2011 and by the Matsumae International Foundation in Japan.
\emph{Centro de Fisica do Porto} is partially funded by the Foundation for 
Science and Technology of Portugal (FCT). The work of V.G. is supported  
by the FCT fellowship SFRH/BD/68313/2010.

\appendix
\section{SO(6) projectors}
\label{App:Projectors}
The $S0(6)$ projectors used in the main text were constructed in \cite{Dolan:2001tt}.
In our notation, they read
\begin{align}
P_1&=\frac{1}{20} y_{12}^4 y_{34}^4 \,,\\
P_{15}&=\frac{1}{4} y_{12}^2 y_{34}^2
\left( y_{24}^2 y_{13}^2-y_{23}^2 y_{14}^2\right),\\
P_{20}&=\frac{1}{10} y_{12}^2 y_{34}^2
\left( 3 y_{24}^2 y_{13}^2+ 3 y_{23}^2 y_{14}^2-
y_{12}^2 y_{34}^2\right),\\
P_{84}&=\frac{1}{3}
\left( y_{13}^4 y_{24}^4+y_{23}^4 y_{14}^4 \right)
+\frac{1}{30} y_{12}^4 y_{34}^4-\frac{2}{3}
y_{13}^2 y_{32}^2y_{24}^2 y_{41}^2-\frac{1}{6}
y_{12}^2 y_{34}^2
\left(  y_{24}^2 y_{13}^2+  y_{23}^2 y_{14}^2\right),\\
P_{105}&=\frac{1}{6}
\left( y_{13}^4 y_{24}^4+y_{23}^4 y_{14}^4 \right)
+\frac{1}{60} y_{12}^4 y_{34}^4+\frac{2}{3}
y_{13}^2 y_{32}^2y_{24}^2 y_{41}^2-\frac{2}{15}
y_{12}^2 y_{34}^2
\left(  y_{24}^2 y_{13}^2+  y_{23}^2 y_{14}^2\right),\\
P_{175}&=\frac{1}{2}
\left( y_{13}^4 y_{24}^4-y_{23}^4 y_{24}^4 \right)
-\frac{1}{4}
y_{12}^2 y_{34}^2
\left(  y_{24}^2 y_{13}^2-  y_{23}^2 y_{14}^2\right).
\end{align}
Comparing with (\ref{G1234decomp}), we conclude that  
 \begin{align}
A_1&=1+ \frac{u^2(1+v^2)}{20v^2} +
   \frac{  u (u+10 (v+1))}{15  v(N^2-1)} +\frac{  2u \left(u^2-8 u (v+1)+10 (v
   (v+4)+1)\right) F(u,v)}{15 v (N^2-1)}\nonumber\,,\\
   A_{15}&=
    \frac{u^2(v^2-1)}{20v^2}
    -\frac{ 2 u (1-v)}{ 5  v(N^2-1)}
    -\frac{ 2 u
   (v-1) (u-2 (v+1)) F(u,v)}{5v(N^2-1)}\nonumber\,,\\
   A_{20}&=
    \frac{u^2(1+v^2 )}{20v^2}
  +\frac{  u (u+10 (v+1))}{30  v(N^2-1)}
    +\frac{  u
   \left(u^2-5 u (v+1)+10 (v-1)^2\right) F(u,v)}{15
   v(N^2-1)} \,,\\
   A_{84}&=
   \frac{u^2(1+v^2 )}{20v^2}
 -\frac{  u^2}{10  v(N^2-1)}
  -\frac{   u^2 (u-3 (v+1)) F(u,v)}{ 5
   v(N^2-1)} \nonumber\,,\\
   A_{105}&=
   \frac{u^2(1+v^2 )}{20v^2}
 +\frac{  u^2}{5  v(N^2-1)}
  +\frac{2  u^3
   F(u,v)}{5v(N^2-1)}\nonumber\,, \\
   A_{175}&=
   \frac{u^2( v^2 -1)}{20v^2}
  +\frac{ 2 u^2 (v-1)
   F(u,v)}{5v(N^2-1)} \nonumber\,.
\end{align}

\section{Anomalous dimensions and  OPE coefficients}
\label{App:GammaOPEcoefs}
For completeness we present expressions for the anomalous dimensions of leading twist operators and associated OPE coefficients   derived in \cite{Eden:2012rr}. 
The perturbative functions $\gamma_n(J)$, as defined in the main text in (\ref{eq:AnomalousDim}), are given by 
\begin{align}
\gamma_1  &=  8  S_1  \,, \\
\gamma_2  &= 8\!\left(2  S_{-2, 1} -  S_{-3} - 2  S_{-2}  S_{1} - 2  S_{1}  S_{2} - S_{3} \right)  , \\
 \gamma_3  &= 64 \big(3  S_{-5} + 8  S_{-4}  S_{1} + 
 S_{-2}^2  S_{1} + 6  S_{-3}  S_{1}^2 + 
 S_{-3}  S_{2} + 4  S_{-2}  S_{1}  S_{2} + 
 2  S_{1}  S_{2}^2+  2  S_{-2}  S_{3} \\
 &+ 
 2  S_{1}^2  S_{3} + S_{2}  S_{3} + 
 3  S_{1}  S_{4} + S_{5} - 6  S_{-4, 1} - 
 12  S_{1}  S_{-3, 1} - 6  S_{-3, 2}-  4  S_{1}^2  S_{-2, 1}- 2  S_{2}  S_{-2, 1} \nonumber \\
 &  - 
 10  S_{1}  S_{-2, 2} - 6  S_{-2, 3} + 
 12  S_{-3, 1, 1} + 16  S_{1}  S_{-2, 1, 1}+   12  S_{-2, 1, 2} + 12  S_{-2, 2, 1} - 24  S_{-2, 1, 1, 1} \big) \,.\nonumber
\end{align}
The definition of the harmonic sums is given in equation  (\ref{Hsums}) of the next appendix,  and we omitted their argument, which is $J$.

The perturbative expansion of the square of OPE coefficients  in (\ref{eq:aJ}) is given by
\begin{align}
a_{0}&=1\,, \ 
 &a_{1}&=-4S_2\,, \\
 a_{2}&=16 \left(3\zeta_3    S_1  +c_{2,4}\right) ,\  
 &a_{3}&=64 \left(\zeta_5c_{3,1}+\zeta_3c_{3,3}+c_{3,6}\right) ,  
\end{align} 
where 
\begin{align} 
c_{2,4} = \ & \frac{5}{2}  S_{-4} + S_{-2}^2 + 2  S_{-3}  S_{1} + 
 S_{-2}  S_{2} + S_{2}^2 + 2  S_{1}  S_{3} + 
 \frac{5}{2}  S_{4} - 2  S_{-3, 1} - S_{-2, 2} - 2  S_{1, 3}  \,, \\
c_{3,1} = \ &  - \frac{25}{2}  S_1  \,, \\
c_{3,3} = \ & -  3  S_{-3} - 10  S_{-2}  S_{1} + \frac{4}{3}  S_{1}^3 - 
 6  S_{1}  S_{2} - \frac{4}{3}  S_{3} + 6  S_{-2, 1}  \,,\\ 
c_{3,6} = \ &  - 11  S_{-6} + \frac{5}{2}  S_{-3}^2 - 5  S_{-4}  S_{-2} - 
 \frac{41}{2}  S_{-5}  S_{1} - S_{-3}  S_{-2}  S_{1} - 
 5  S_{-4}  S_{1}^2 - 2  S_{-2}^2  S_{1}^2 \\ 
 & +  \frac{4}{3}  S_{-3}  S_{1}^3 - \frac{13}{2}  S_{-4}  S_{2} - 
 \frac{3}{2}  S_{-2}^2  S_{2} - 10  S_{-3}  S_{1}  S_{2} - 
 2  S_{-2}  S_{2}^2 - S_{2}^3 - \frac{16}{3}  S_{-3}  S_{3} \nonumber\\
 & -  8  S_{-2}  S_{1}  S_{3} - 6  S_{1}  S_{2}  S_{3} - 
 3  S_{3}^2 - 3  S_{-2}  S_{4} + 9  S_{1}^2  S_{4} - 
 4  S_{2}  S_{4} + \frac{15}{2}  S_{1}  S_{5} - \frac{13}{2}  S_{6} \nonumber \\
 & +  14  S_{-5, 1} + 11  S_{1}  S_{-4, 1} + 9  S_{-4, 2} - 
 12  S_{1}  S_{-3, -2} + 10  S_{-2}  S_{-3, 1} - 4  S_{1}^2  S_{-3, 1} \nonumber \\
 & + 8  S_{2}  S_{-3, 1} + 
 4  S_{1}  S_{-3, 2} + 9  S_{-3, 3} - 
 10  S_{-3}  S_{-2, 1} + 
 14  S_{-2}  S_{1}  S_{-2, 1} - 
 \frac{8}{3}  S_{1}^3  S_{-2, 1} \nonumber \\
 & +  4  S_{1}  S_{2}  S_{-2, 1} + \frac{20}{3}  S_{3}  S_{-2, 1} + 
 10  S_{-2, 1}^2 + 10  S_{-2}  S_{-2, 2} - 
 6  S_{1}^2  S_{-2, 2} + 6  S_{2}  S_{-2, 2} \nonumber \\
 & +  6  S_{1}  S_{-2, 3} + 11  S_{-2, 4} - 
 6  S_{2}  S_{1, 3} - 4  S_{1}  S_{1, 4} - 
 4  S_{1, 5} + 4  S_{1}  S_{2, 3} + 4  S_{2, 4} - 
 12  S_{-4, 1, 1} \nonumber \\
 & +  8  S_{1}  S_{-3, 1, 1} - 
 2  S_{-3, 1, 2} - 2  S_{-3, 2, 1} - 
 24  S_{1}  S_{-2, -2, 1} - 20  S_{-2}  S_{-2, 1, 1} + 
 16  S_{1}^2  S_{-2, 1, 1}  \nonumber \\
 & -  8  S_{2}  S_{-2, 1, 1} + 
 16  S_{1}  S_{-2, 1, 2} - 6  S_{-2, 1, 3} + 
 16  S_{1}  S_{-2, 2, 1} + 4  S_{-2, 2, 2} - 
 6  S_{-2, 3, 1} - 4  S_{1}  S_{1, 1, 3} \nonumber \\
 & -  8  S_{1, 1, 4} + 8  S_{1, 3, 2} - 8  S_{-3, 1, 1, 1} - 
 48  S_{1}  S_{-2, 1, 1, 1} - 20  S_{-2, 1, 1, 2} - 
 20  S_{-2, 1, 2, 1} - 20  S_{-2, 2, 1, 1} \nonumber \\
 & + 16  S_{1, 1, 1, 3} + 64  S_{-2, 1, 1, 1, 1} \,.\nonumber
\end{align}

\section{Harmonic sums and reflection symmetry}
\label{App:Harmonics}

Harmonic sums can be recursively defined as,
\begin{equation}
S_{a_1,a_2,\dots,a_n}(x)=\sum_{y=1}^{x}\frac{({\rm sign}(a_{1}))^{y}}{y^{|a_1|}} \, S_{a_2,\dots,a_n}(y)\,, 
\label{Hsums}
\end{equation}
where the term with no indices is defined as $S\!\left(y\right)=1$ and the argument $x$ is assumed to be an integer. However, harmonic sums can be analytically continued to every value of $x$, details can be seen in \cite{Kotikov:2005gr,Blumlein:2000hw}.

There are certain quantities in the Regge kinematics that can be written as an antisymmetric or symmetric combination of harmonic sums at $x$ and $-1-x$, where $x=(i\nu-1)/2$.  For example, the BFKL spin is symmetric under the exchange of $x\rightarrow -1-x$ while the pre-factor $\alpha$ contains an antisymmetric factor under this symmetry. In the perturbative regime the coefficients of the $\ln\sigma$ terms in the four-point function are written as products like $j_k^n \alpha_q^m$ which have well defined symmetry under $x\rightarrow -1-x$, but involve products of harmonic sums with arguments $x$ and $-x-1$. It is sometimes possible to express these functions in terms of linear combinations of harmonic sums without mixed arguments. As a simple example consider, 
\begin{align}
S_1(x) S_2(-1 - x) + S_1(-1 - x) S_2(x)
\end{align}
which is equivalent to, 
\begin{align}
6 \zeta_3 + \big(S_1(x) S_2(x) + S_3(x) -  2S_{2,1}(x)+(x\leftrightarrow -1-x)\big).
 \label{eq:newsymmetryrel}
\end{align}
Identities like this one are sometimes useful to rewrite expressions in a simpler form as in equation (\ref{eq:alpha3simple}). They can also be used in more technical aspects such as finding expansions near some point.
We have discovered these identities performing numerical experiments but we are not aware of analytic derivations.

\section{Integral representation for leading logs}
\label{app:IntRep}

Consider the sum
\be
T_n(z)=
\sum_{J=0 \atop {\rm even} }^\infty
\frac{J!^2}{(2J)!} \, \left(S_1(J) \right)^n
z^J
  F \left( 2+2J,z\right) .
\ee
It is not hard to show that 
\be
T_0(z)=\frac{2-z}{2(1-z)}\,.
\ee
Let us use the integral representation of the hypergeometric function
\be 
\frac{J!^2}{(2J)!}  
 \,z^J  F \left( 2+2J,z\right) =
 (2J+1)\int_0^1\frac{dt}{1-zt}
 \left(\frac{zt(1-t)}{1-zt}
 \right)^J  ,
\ee
and of the harmonic sum
\be
S_1(J)=\int_0^1dx \frac{1-x^J}{1-x}\,,
\ee
to write
\be
T_n(z)=\int_0^1\frac{dt}{1-zt}\int_0^1 \prod_{k=1}^n \frac{dx_k}{1-x_k}
\sum_{J=0 \atop {\rm even} }^\infty
(2J+1) 
 \left(\frac{zt(1-t)}{1-zt}
 \right)^J
 \prod_{i=1}^n\left(1-x_i^J\right) .
\ee
We can now expand the product
\be
 \prod_{i=1}^n \left(1-x_i^J\right)=
 \sum_{\{w\}} (-1)^{\sum_i w_i} \left(\prod_{j=1}^n x_j^{w_j}\right)^J ,
\ee
where the sum runs over all lists $\{w\}=\{w_1,\dots,\w_n\}$ with $w_i=0,1$. 
Finally, we can use
\be
\sum_{J=0 \atop {\rm even} }^\infty
(2J+1) 
 y^J
=\frac{1+3y^2}{(1-y^2)^2}\,,
\ee
to write
\be
T_n(z)=\int_0^1\frac{dt}{1-zt}\int_0^1 \prod_{k=1}^n \frac{dx_k}{1-x_k}
\sum_{\{w\}} (-1)^{\sum_i w_i} 
\frac{1+3y^2}{(1-y^2)^2} \,,
\label{intrepLLOG}
\ee
where
\be
y=\frac{zt(1-t)}{1-zt}\prod_{j=1}^n x_j^{w_j}\,.
\ee

\section{Lorentzian OPE limit at four loops}
\label{ap:4loopsOPE}
In this appendix we give the explicit form of our predictions for the behaviour of the four point function at four loops in the Lorentzian OPE limit. 
The results are given in terms of harmonic polylogarithms following the notation of (\ref{kappa}):
\begin{align}
&\k_1^{(4)}=\frac{8}{3}(10
   H_{1,4}-20
   H_{5}+10
   H_{2,3}+14
   H_{3,2}+6
   H_{4,1}-6
   H_{1,3,1}+3
   H_{2,1,2}-3
   H_{2,2,1}\nonumber\\
&+6
   H_{3,1,1}-3
   H_{1,2,1,1}+3
   H_{2,1,1,1})\,,
\\\nonumber
\\
&\tilde{\k}_1^{(4)}= -\frac{8}{3}(36 H_{1,4}+36
   H_{2,3}+44
   H_{3,2}+24
   H_{4,1}+24
   H_{1,1,3}+28
   H_{1,2,2}+16
   H_{1,3,1}\nonumber\\
&+28
   H_{2,1,2}+16
   H_{2,2,1}+28
   H_{3,1,1}+15
   H_{1,1,1,2}+15
   H_{1,1,2,1}+12
   H_{1,2,1,1}+18
   H_{2,1,1,1})\,,
\\\nonumber
\\
&\k_2^{(4)}=8(60 H_{6}-10
   H_{1,5}-10
   H_{2,4}-22
   H_{3,3}-40
   H_{4,2}-38
   H_{5,1}+8
   H_{1,1,4}+8
   H_{1,2,3}\nonumber\\
&-2
   H_{1,3,2}+16
   H_{1,4,1}+7
   H_{2,1,3}-3
   H_{2,2,2}-8
   H_{3,1,2}-2
   H_{3,2,1}-24
   H_{4,1,1}+8
   H_{1,1,1,3}\nonumber\\
&+
   H_{1,2,1,2}+8
   H_{1,2,2,1}-2
   H_{1,3,1,1}+3
   H_{2,1,1,2}-4
   H_{2,1,2,1}-8
   H_{2,2,1,1}-6
   H_{3,1,1,1}\nonumber\\
&+4
   H_{1,1,1,1,2}-4
   H_{1,2,1,1,1})+1152\zeta_3 ( 2H_{3}-
   H_{1,2})\,,
\end{align}
\begin{align}
&\tilde{\k}_2^{(4)}=8(60 H_{1,5}+60
   H_{2,4}+84
   H_{3,3}+120
   H_{4,2}+96
   H_{5,1}+24
   H_{1,1,4}+36
   H_{1,2,3}\nonumber\\
&+50
   H_{1,3,2}+30
   H_{1,4,1}+36
   H_{2,1,3}+50
   H_{2,2,2}+30
   H_{2,3,1}+58
   H_{3,1,2}+34
   H_{3,2,1}\nonumber\\
&+72
   H_{4,1,1}+13
   H_{1,1,1,3}+19
   H_{1,1,2,2}+14
   H_{1,1,3,1}+23
   H_{1,2,1,2}+3
   H_{1,2,2,1}\nonumber\\
&+18
   H_{1,3,1,1}+23
   H_{2,1,1,2}+19
   H_{2,1,2,1}+22
   H_{2,2,1,1}+26
   H_{3,1,1,1}+8
   H_{1,1,1,1,2}\nonumber\\
&-6
   H_{1,1,1,2,1}-4
   H_{1,1,2,1,1}+2
   H_{1,2,1,1,1})+1152\zeta_3 (4 H_{1,2}+2
   H_{2,1}+3H_{1,1,1})\,,
\\\nonumber
\\
&\k_{3}^{(4)}=-8 (280 H_{7}+20
   H_{1,6}+20 H_{2,5}-4
   H_{3,4}-88
   H_{4,3}-208
   H_{5,2}-300
   H_{6,1}+80
   H_{1,1,5}+64
   H_{1,2,4}\nonumber\\
&+28
   H_{1,3,3}-14
   H_{1,4,2}+22
   H_{1,5,1}+74
   H_{2,1,4}+26
   H_{2,2,3}-6
   H_{2,3,2}-54
   H_{2,4,1}+36
   H_{3,1,3}-20
   H_{3,2,2}\nonumber\\
&-36
   H_{3,3,1}-38
   H_{4,1,2}-42
   H_{4,2,1}-120
   H_{5,1,1}+46
   H_{1,1,1,4}+26
   H_{1,1,2,3}+18
   H_{1,1,3,2}-10
   H_{1,1,4,1}\nonumber\\
&+18
   H_{1,2,1,3}+11
   H_{1,2,2,2}+5
   H_{1,2,3,1}+7
   H_{1,3,1,2}+11
   H_{1,3,2,1}-42
   H_{1,4,1,1}+18
   H_{2,1,1,3}+15
   H_{2,1,2,2}\nonumber\\
&-25
   H_{2,1,3,1}+2
   H_{2,2,1,2}-12
   H_{2,2,2,1}-28
   H_{2,3,1,1}-H_{3,1,1,2}-13 H_{3,1,2,1}-32
   H_{3,2,1,1}-14
   H_{4,1,1,1}\nonumber\\
&+12
   H_{1,1,1,1,3}+22
   H_{1,1,1,2,2}-8
   H_{1,1,1,3,1}+11
   H_{1,1,2,1,2}+5
   H_{1,1,2,2,1}-2
   H_{1,1,3,1,1}+11
   H_{1,2,1,1,2}\nonumber\\
&-13
   H_{1,2,1,2,1}-26
   H_{1,2,2,1,1}-12
   H_{1,3,1,1,1}-6
   H_{2,1,1,1,2}+2
   H_{2,1,1,2,1}+4
   H_{2,2,1,1,1})-768\zeta_3 (
   H_{1,3}\nonumber\\
&+8 H_{4}-13
   H_{2,2}-20
   H_{3,1}-6
   H_{1,1,2}+4
   H_{1,2,1}-16
   H_{2,1,1}-6
   H_{1,1,1,1})-21120 \zeta_5 H_{2}\,,
\end{align}
\begin{align}
&\tilde{\k}_{3}^{(4)}=-8 (120 H_{1,6}+120
   H_{2,5}+168
   H_{3,4}+336
   H_{4,3}+576
   H_{5,2}+640
   H_{6,1}+24
   H_{1,2,4}\nonumber\\
&+84
   H_{1,3,3}+168
   H_{1,4,2}+164
   H_{1,5,1}+24
   H_{2,1,4}+84
   H_{2,2,3}+168
   H_{2,3,2}+164
   H_{2,4,1}+84
   H_{3,1,3}\nonumber\\
&+168
   H_{3,2,2}+140
   H_{3,3,1}+216
   H_{4,1,2}+152
   H_{4,2,1}+320
   H_{5,1,1}+14
   H_{1,1,1,4}+6
   H_{1,1,2,3}+44
   H_{1,1,3,2}\nonumber\\
&+36
   H_{1,1,4,1}+18
   H_{1,2,1,3}+30
   H_{1,2,2,2}+16
   H_{1,2,3,1}+46
   H_{1,3,1,2}+10
   H_{1,3,2,1}+80
   H_{1,4,1,1}+30
   H_{2,1,1,3}\nonumber\\
&+44
   H_{2,1,2,2}+46
   H_{2,1,3,1}+66
   H_{2,2,1,2}+14
   H_{2,2,2,1}+80
   H_{2,3,1,1}+72
   H_{3,1,1,2}+36
   H_{3,1,2,1}+80
   H_{3,2,1,1}\nonumber\\
&+72
   H_{4,1,1,1}+12
   H_{1,1,1,1,3}+8
   H_{1,1,1,2,2}-14
   H_{1,1,1,3,1}+H_{1,1,2,1,2}-25 H_{1,1,2,2,1}-2
   H_{1,1,3,1,1}\nonumber\\
&+3
   H_{1,2,1,1,2}-15
   H_{1,2,1,2,1}+22
   H_{1,2,2,1,1}+10
   H_{1,3,1,1,1}+26
   H_{2,1,1,1,2}-24
   H_{2,1,1,2,1}-8
   H_{2,1,2,1,1}\nonumber\\
&+6
   H_{2,2,1,1,1}-8
   H_{1,1,1,1,2,1}+4
   H_{1,1,1,2,1,1}+4
   H_{1,1,2,1,1,1})-768\zeta_3 (28 H_{2,2}+36
   H_{3,1}\nonumber\\
&+2
   H_{1,1,2}+3
   H_{1,2,1}+23
   H_{2,1,1}-2
   H_{1,1,1,1})-21120 \zeta_5 H_{1,1}\,.
\end{align}

\section{Recursion relation for leading logs}
\label{App:LLOG}

\label{App:LLOGREGGE}
Consider the integral
\begin{align}
I_n &=\int d\nu 
 \frac{\tanh\!\left(\frac{\pi\n}{2}\right)}
 {\cosh^2\!\left(\frac{\pi\n}{2}\right)} 
 \left[S_1\left(\frac{i\n-1}{2}\right)
 +S_1\left(\frac{-i\n-1}{2}\right)
 \right]^n
 \sin(\nu\rho)\,.
\end{align}
For $n=0$ the integral gives
\be
I_0=\frac{4\r^2}{\pi^2\sinh \r}\,.
\ee
When $n>0$ we use the following integral representation of the harmonic sum
\begin{align}
S_1\left(\frac{i\n-1}{2}\right)
 +S_1\left(\frac{-i\n-1}{2}\right)
 &=\int_0^1 dx \frac{2-x^{-\frac{1+i\n}{2}}-x^{-\frac{1-i\n}{2}}}{1-x}\\
 &=\int_0^1 \frac{2dx}{1-x}\left[1-\frac{1}{\sqrt{x }}\cos\left(\frac{\n}{2}\log x\right)\right] .
\end{align}
In fact, we need to use this $n$-times
\begin{align}
\left[S_1\left(\frac{i\n-1}{2}\right)
 +S_1\left(\frac{-i\n-1}{2}\right)
 \right]^n
  =\int_0^1 \prod_{k=1}^n\frac{2dx_k}{1-x_k}
  \prod_{k=1}^n\left[1-\frac{1}{\sqrt{x_k}}\cos\left(\frac{\n}{2}\log x_k \right)\right] .
\end{align}
The next step is to use the following identity 
\begin{align}
 \sin(\nu\rho)
  \prod_{k=1}^n\left[1-\frac{1}{\sqrt{x_k}}\cos\left(\frac{\n}{2}\log x_k \right)\right]
  =\sum_{\{w\}} \frac{\sin(\nu\rho(\{x\},\{w\}))}
  {\prod_{k=1}^n (-2\sqrt{x_k})^{|w_k|}}  \,,
\end{align}
where  the sum runs over all lists $\{w\}=\{w_1,\dots,\w_n\}$ with $w_i=-1,0,1$ and 
\be
\rho(\{x\},\{w\}) =\r +\sum_{i  \in  S}  \frac{w_i}{2}\log x_i\,.
\ee
This identity can be easily derived using the exponential representation of the trigonometric functions involved.

Putting everything together, we can write
\begin{align}
I_n&= 
\int_0^1 \prod_{k=1}^n\frac{2dx_k}{1-x_k}
\sum_{\{w\}} \frac{1}{\prod_{i=1}^n (-2\sqrt{x_i})^{|w_i|}} 
\int d\nu 
 \frac{\tanh\left(\frac{\pi\n}{2}\right)}
 {\cosh^2\left(\frac{\pi\n}{2}\right)} 
 \sin\Big(\nu \rho(\{x\},\{w\})  \Big)\\
&=\frac{4 }{\pi^2 }
\int_0^1 \prod_{k=1}^n\frac{2dx_k}{1-x_k}
\sum_{\{w\}} \frac{1}{\prod_{i=1}^n (-2\sqrt{x_i})^{|w_i|}} 
\frac{ \Big( \rho(\{x\},\{w\}) \Big)^2}{ \sinh \rho(\{x\},\{w\}) }\,,
\end{align}
where the integral over $\nu$ became exactly of the same form of the one in $I_0$.
From the last expression, one easily deduces the following recursion relation
\be
I_n(\rho) = \int_0^1 \frac{2 dx_n}{1-x_n} \left[
I_{n-1}(\rho)-\frac{1}{2\sqrt{x_n}} I_{n-1}\left(\rho+\half \log x_n\right)
-\frac{1}{2\sqrt{x_n}}I_{n-1}\left(\rho-\half \log x_n\right)
\right],
\nonumber
\ee
where the three terms correspond to $w_n=0,$ 1 and $-1$, respectively.
 
Finally, it is convenient to use the variable $r=e^{-2\r}$ and define
$I_n(\r)=\frac{2}{\pi^2 }e^{-\r} B_n(e^{-2\r})$. Then, the recursion relation simplifies to
\be
B_n(r) = \int_0^1 \frac{ dx }{1-x } \left[
2B_{n-1}(r)-  B_{n-1}\left(rx\right)
-  \frac{1}{x}B_{n-1}\left(\frac{r}{ x}\right)
\right] ,
\ee
which can be iterated starting from 
\be
B_0(r)= \frac{\log^2r}{1-r}\,.
\ee

\section{Analytic continuation of the four-point function}
\label{AnalyticContinuation}

Here we give some details on the analytic continuation of the four-point function required to analyse the Regge limit described in Section \ref{Reggeweakcoupling}. The two-loop calculation was performed in \cite{ourBFKL} and the only contribution came from the $[\phi^{(1)}(z,\bar z)]^2$ contribution to $g$ given in equation (\ref{fgh2loops}). At three loops we recall from  (\ref{Reggelimform}) that we need to consider the contributions to the functions $g$ and $h$. In fact all contributions to $h$ at three loops (from the Easy function $E$ and the Hard function $H^{(b)}$) are given in terms of single-valued combinations of harmonic polylogarithms \cite{Drummond:2013nda} and calculating the analytic continuation is straightforward. The result is that at three loops the function $h$ gives no contribution in the Regge limit (i.e. it is power suppressed in the limit $\sigma \rightarrow 0$).

The contributions to the function $g$ are of two types. Firstly there are terms of the form $\phi^{(1)} \phi^{(2)}$ coming from the product of one-loop and two-loop ladder integrals. These ladder contributions are again given in terms of single-valued combinations of harmonic polylogarithms and again their analytic continuation is straightforwardly obtained. The resulting contribution to the analytically continued four-point function in the Regge limit is,
\be
{\rm ladders} \rightarrow 2 \pi^2 \frac{r}{(1-r)^2} \log^2 r  (2 \pi^2 + \log^2 r  - 4 \log^2 \sigma - 4 i \pi \log\sigma)\,.
\label{ladders}
\ee

The second type of contribution to $g$ comes from the Hard function $H^{(a)}$ which is given in terms of a single-valued combination of two-variable multiple polylogarithm functions \cite{Drummond:2013nda}. In general these functions are specified by a weight vector $w = a_1 a_2 \ldots a_n $. If $w$ is just a string of zeros, we define $G(0_n;x) = \frac{1}{n!} \log^n x$. Then, if we write a general weight vector $w$ as $w = a_1 w'$ with $w' = a_2 \dots a_n$, the remaining functions can be defined recursively via
\be
G(w;x) = G(a_1,a_2,\ldots , a_n;x) = G(a_1,w';x)  = \int_0^x \frac{dt}{t-a_1} G(w';t)\,. 
\label{Gdef}
\ee
Such multiple polylogarithms obey a shuffle product relation,
\be
G(w_1;z)G(w_2;z) = G(w_1 \sha w_2;z)\,,
\label{Gshuffle}
\ee
and, if the word $w$ does not have trailing zeros, a rescaling relation,
\be
G(a_1,\ldots,a_n; x) = G(\lambda a_1, \ldots, \lambda a_n ; \lambda x)\,, \qquad a_n \neq 0\,.
\label{Grescaling}
\ee
The harmonic polylogarithms used throughout are special cases where the weight vector consists only of zeros and ones. Due to unfortunate choices of conventions the precise relation involves a sign,
\be
G(w;x) = (-1)^{d(w)} H(w;x)\,, \qquad a_i \in \{0,1\}\,,
\label{GtoHPL}
\ee
where $d(w)$ is the number of ``1''  entries in $w$.

Obtaining the analytic continuation for the contributions to $g$ coming from the Hard function $H^{(a)}$ is slightly more involved. Here we describe a method for obtaining the analytic continuation based on an integral formula for $H^{(a)}$ given in Appendix B of \cite{Drummond:2013nda},
\begin{align}
&H^{(a)}(1-z,1-\bar z) =\notag \\
&\qquad= \bigl(2 H_{0,0}(u) + 4 H_{0}(u)H_{0}(v) + 8 H_{0,0}(v)\bigr) \bigl(\mathcal{L}_{0,0,1,1} +  \mathcal{L}_{1,1,0,0} -  \mathcal{L}_{0,1,1,0} - \mathcal{L}_{1,0,0,1}\bigr) \notag \\
& \qquad \quad - 8F_5 \bigl(H_0(u) + 2 H_0(v)\bigr) + F_6\,.
\label{Hsdef}
\end{align}
In the above equation the $\mathcal{L}$ functions are single-valued combinations of harmonic polylogarithms, defined by Brown \cite{fbsvp}. We refer the reader to \cite{Drummond:2013nda} for all the conventions on single-valued polylogarithms. The functions $F_5$ and $F_6$ are given by integral formulae,
\begin{align}
\label{Fnintform}
F_n(z,\bar z) &= \int dt \biggl[ \frac{X_{n-1}(t,\bar z)}{t} - \frac{Y_{n-1}(t,\bar z)}{1-t} + \frac{Z_{n-1}(t,\bar z)}{t-\bar z}\biggr]  \\
&= F_n^X(z,\bar z) + F_n^Y(z,\bar z) + F_n^Z(z,\bar z)\,.
\end{align}
For $F_5$ the integrand in (\ref{Fnintform}) is given by the following three functions which are again single-valued combinations of harmonic polylogarithms,
\begin{align}
X_4(x,\bar x) &= (\mathcal{L}_{0,0,1,1} - \mathcal{L}_{1,1,0,0} - \mathcal{L}_{0,1,1,1}+\mathcal{L}_{1,1,1,0})\,, \notag \\
Y_4(x,\bar x) &= (\mathcal{L}_{0,0,0,1} - \mathcal{L}_{1,0,0,0} - \mathcal{L}_{0,0,1,1} + \mathcal{L}_{1,1,0,0})\,, \notag \\
Z_4(x,\bar x) &= (\mathcal{L}_{0,0,1,1} + \mathcal{L}_{1,1,0,0} - \mathcal{L}_{0,1,1,0} - \mathcal{L}_{1,0,0,1})\,.
\end{align}
For $F_6$ the integrand in  (\ref{Fnintform}) is given by the following three functions,\
\begin{align}
X_5 &= 20 \mathcal{L}_{0,0,0,1,1} + 12 \mathcal{L}_{0,0,1,1,0} - 32 \mathcal{L}_{0,0,1,1,1} - 8 \mathcal{L}_{0,1,0,1,1} - 12 \mathcal{L}_{0,1,1,0,0} - 8 \mathcal{L}_{0,1,1,0,1} \notag \\
&+16 \mathcal{L}_{0,1,1,1,1} - 8\mathcal{L}_{1,0,0,1,1} + 8 \mathcal{L}_{1,0,1,1,0} - 20 \mathcal{L}_{1,1,0,0,0} + 8 \mathcal{L}_{1,1,0,0,1} + 8 \mathcal{L}_{1,1,0,1,0} \notag \\
&+ 32 \mathcal{L}_{1,1,1,0,0} - 16 \mathcal{L}_{1,1,1,1,0} - 16\mathcal{L}_{1,1} \zeta_3\,, \\
Y_5 &= 20 \mathcal{L}_{0,0,0,0,1} - 32 \mathcal{L}_{0,0,0,1,1} - 8 \mathcal{L}_{0,0,1,1,0} + 16 \mathcal{L}_{0,0,1,1,1} - 8 \mathcal{L}_{0,1,0,0,1} + \mathcal{L}_{0,1,1,0,0} \notag \\
&- 20 \mathcal{L}_{1,0,0,0,0} + 8 \mathcal{L}_{1,0,0,1,0} + 16 \mathcal{L}_{1,0,0,1,1} + 8 \mathcal{L}_{1,0,1,0,0} + 32 \mathcal{L}_{1,1,0,0,0} - 16 \mathcal{L}_{1,1,0,0,1}  \notag \\
&- 16 \mathcal{L}_{1,1,1,0,0} - 16 \mathcal{L}_{1,0} \zeta_3 + 64 \mathcal{L}_{1,1} \zeta_3\,. \\
Z_5 &= 32F_5\,.
\end{align}

In the case of both $F_5$ and $F_6$ the $X$ and $Y$ parts of the integrand can be integrated immediately to obtain $F_n^X$ and $F_n^Y$ in terms of harmonic polylogarithms. The analytic continuation of these terms can therefore be obtained easily. For the $F_n^Z$ contributions we use the fact that the discontinuity around $z=1$ can be moved through the integral sign,
\be
\Delta_{1-z} F_n^Z(z,\bar z) = \int_1^z \frac{dt}{t-\bar z} \Delta_{1-t} Z_{n-1}(t,\bar z)\,.
\ee
Since $Z_4$ is a combination of harmonic polylogarithms its discontinuity can be easily calculated. Then to obtain $\Delta_{1-z}F_5$ it remains to perform the integral. This can be done using multiple (or Goncharov) polylogarithms (see equations (\ref{Gdef}) and (\ref{GtoHPL})),
\be
\int_0^z \frac{dt}{t-\bar z} H_w(t) = (-1)^{d(w)}G(\bar z, w;z)\,.
\ee
Having analytically continued around $z=1$ (there is no contribution from $z=0$) it remains to take the Regge limit $z,\bar z \rightarrow 0$ with fixed ratio $r= z/\bar z$. This can be done by first extracting any trailing zeros from the word $w$ appearing in the $G$-functions above by using the shuffle relations (\ref{Gshuffle}). Having made any $\log z$ and $\log \bar z$ explicit, one may rescale the arguments of the $G$-functions using (\ref{Grescaling}),
\be
G(\bar z, w;z) = G(1,w/\bar z ; r).
\ee
Then any $G$-functions exhibiting a letter $1/\bar z$ in the weight vector will be power suppressed in the Regge limit and may be dropped. Finally one obtains an expression in terms of powers of $H_0(\sigma) = \log \sigma$ and harmonic polylogarithms of argument $r$. The result for the Regge limit of $F_5$ is 
\be
F_5 \rightarrow 32 \pi^2 \bigl( - 2 H_{0}(\sigma) H_{0, 0}(r) + 
  H_{0, 0, 0}(r) + 2 H_{1, 0, 0}(r) - 2 \zeta_3+i \pi H_{0, 0}(r)\bigr)\,.
\ee

For $F_6$ we can perform the same analysis as above with the only difference that $Z_5 = 32 F_5$ so that we need to recycle our previous result for $\Delta_{1-z} F_5(z,\bar z)$ when calculating $\Delta_{1-z} F_6^Z(z,\bar z)$. This leads to some terms with two $\bar z$ appearing in the arguments of the $G$ functions. Otherwise the analysis is very similar and we find that the Regge limit of $F_6$ is given by
\begin{align}
F_6 \rightarrow 8 \pi^2 \bigl( & - 8 H_{0, 0}(r) H_{0, 0}(\sigma)  + 4 H_{0}(\sigma) H_{0, 0, 0}(r) + 8 H_{0}(\sigma) H_{1, 0, 0}(r) - 4 H_{2, 0, 0}(r) - 
   H_{0, 0, 0, 0}(r) \notag \\
   & - 4 H_{1, 0, 0, 0}(r) - 8 H_{1, 1, 0, 0}(r) + 4 H_{0, 0}(r) \zeta_2 + 6 \zeta_4  + 4 H_{0}(r) \zeta(3) - 8 H_{0}(\sigma) \zeta_3 \notag \\
   &+ 8 H_{1}(r) \zeta_3+4 i \pi H_{0}(\sigma) H_{0, 0}(r) - 2 i \pi H_{0, 0, 0}(r) - 4 i \pi H_{1, 0, 0}(r) + 4 i \pi \zeta_3 \bigr) \,.
\end{align}
Combining the above calculations we can now obtain the Regge limit of $H^{(a)}(1-z,1-\bar z)$,
\begin{align}
H^{(a)}(1-z,1-\bar z) \rightarrow 8 \pi^2 \bigl(& - 2 H_{0, 0}(r) H_{0, 0}(\sigma)  + 2 H_{0}(\sigma) H_{0, 0, 0}(r)  + 4 H_{0}(\sigma) H_{1, 0, 0}(r) \notag \\
&- 4 H_{2, 0, 0}(r) - H_{0, 0, 0, 0}(r) - 4 H_{1, 0, 0, 0}(r) - 8 H_{1, 1, 0, 0}(r) + 4 H_{0, 0}(r) \zeta_2   \notag \\
&+ 6 \zeta_4 + 4 H_{0}(r) \zeta_3 - 4 H_{0}(\sigma) \zeta_3 + 8 H_{1}(r) \zeta_3 -2 i \pi H_{0}(\sigma) H_{0, 0}(r) \notag \\
&+ 2 i \pi H_{0, 0, 0}(r) + 4 i \pi H_{1, 0, 0}(r) - 4 i \pi \zeta_3\bigr)
\label{Haomzlim}
\end{align}

Similar calculations yield results for $H^{(a)}(z,\bar z)$ and $H^{(a)}\bigl(\tfrac{1}{z},\tfrac{1}{\bar z}\bigr)$,
\begin{align}
\label{Halim}
H^{(a)}(z,\bar z)\, \rightarrow 16 \pi^2 \bigl(&-2 H_{2, 0, 0}(r) - 2 H_{1, 0, 0, 0}(r) - 4 H_{1, 1, 0, 0}(r) + 3 \zeta_4 + 2 H_{0}(r) \zeta_3 + 4 H_{1}(r) \zeta_3\bigr)  \,,
\\
H^{(a)}\bigl( \tfrac{1}{z},\tfrac{1}{\bar z}\bigr) \rightarrow \,\, 8 \pi^2 \bigl(&-2 H_{0, 0}(r) H_{0, 0}(s) + 2 H_{0}(s) H_{0, 0, 0}(r) + 4 H_{0}(s) H_{1, 0, 0}(r) - 4 H_{2, 0, 0}(r) \notag \\
 & - H_{0, 0, 0, 0}(r) - 4 H_{1, 0, 0, 0}(r) - 8 H_{1, 1, 0, 0}(r) - 2 H_{0, 0}(r) \zeta_2] + 6 \zeta_4 + 4 H_{0}(r) \zeta_3 \notag \\
& - 4 H_{0}(s) \zeta_3 + 8 H_{1}(r) \zeta_3\bigr)
\,.
\end{align}

Finally, as dictated by equations (\ref{fgh3loops}) and (\ref{Reggelimform}), we need to take the combination 
\be
\frac{r}{(1-r)^2}\Bigl(-2H^{(a)}(z,\bar z) - H^{(a)}(1-z,1-\bar z) - H^{(a)}\bigl(\tfrac{1}{z},\tfrac{1}{\bar z}\bigr)\Bigr)\,,
\ee
and combine with the ladder contributions (\ref{ladders}) to obtain the results quoted in equations (\ref{xi10}) and (\ref{xi11}) of the main text. Note that the $\log^2 \sigma$ divergence from the ladder contribution (\ref{ladders}) is cancelled by similar divergences in (\ref{Haomzlim}) and (\ref{Halim}) (recall $2H_{0,0}(\sigma) =  \log^2 \sigma$). This is necessary since the leading divergence in the four-point function at three loops should only be a single power of $\log \sigma$. Likewise the imaginary divergent contribution from (\ref{ladders}) is cancelled by a similar contribution from (\ref{Haomzlim}) which is necessary for the coefficient of the leading $\log \sigma$ divergence to be real.

\bibliographystyle{./utphys}
\bibliography{./mybib}

\end{document}